\documentclass[twocolumn]{autart}    
\usepackage{graphicx}          % Include this line if 

\usepackage{amsmath,amssymb,amsfonts}
\usepackage{booktabs}
\usepackage[nolist, nohyperlinks]{acronym}

\usepackage[dvipsnames]{xcolor}
\usepackage{bm}
\usepackage{tabularx}
\usepackage{array}
\newcolumntype{L}[1]{>{\raggedright\arraybackslash}p{#1}}

\acrodef{GP}{Gaussian Process}
\acrodef{SDE}{stochastic differential equation}
\acrodef{STGPKF}{spatiotemporal Gaussian process Kalman filter}
\acrodef{KF}{Kalman filter}
\acrodef{CARE}{Continuous-time algebraic Riccati equation}
\acrodef{AEC}{averaged expected clarity}
\acrodef{AIM}{averaged information matrix}

\newcommand{\reals}{\mathbb{R}}
\newcommand{\R}{\reals}

\renewcommand{\S}{\mathbb{S}}

\newcommand{\pd}{\mathbb{S}_{++}}
\newcommand{\psd}{\mathbb{S}_{+}}

\newcommand{\E}{\mathbb{E}}

\newcommand{\Lcal}{\mathcal{L}}

\newcommand{\Ncal}{\mathcal{N}}

\newcommand{\Rcal}{\mathcal{R}}

\newcommand{\eqn}[1]{\begin{align} #1 \end{align}}
\newcommand{\eqnN}[1]{\begin{align*} #1 \end{align*}}
\newcommand{\seqn}[2][]{
\begin{subequations}
    #1
\begin{align} #2 \end{align}
\end{subequations}
}
\newcommand{\eqnOne}[1]{%
\begin{equation}
\begin{aligned}
#1
\end{aligned}
\end{equation}
}

\newcommand{\bmat}[1]{\begin{bmatrix}#1\end{bmatrix}}

\newenvironment{editone}{\color{blue}}{}

\newcommand{\Ric}{\Rcal}

\renewenvironment{editone}{\color{black}}{}
\newcommand{\codelink}[1]{%
  \leavevmode
  \pdfstartlink attr{/Border [0 0 0]} user{%
    /Subtype /Link
    /A << /S /URI /URI (https://github.com/Lawliet9666/multi-sensor-network-design) >>%
  }#1\pdfendlink}

\begin{document}

\begin{frontmatter}

% \title{Continuous-Time Steady-State Analysis of Multi-Agent Estimation in Spatiotemporal Fields
% $% \thanksref{footnoteinfo}
% $} % Title, preferably not more 
% \title{Fundamental Performance Limits in Multiagent Spatiotemporal Estimation}
% \title{Fundamental Limits of Continuous-Time Multiagent Spatiotemporal Estimation under Randomized Sensing}
\title{Kalman-Bucy Filtering with Randomized Sensing: Fundamental Limits and Sensor Network Design for Field Estimation}
\author[Rob,equalcontribution]{Xinyi Wang\corauthref{correspondingauthor}}\ead{xinyiwang@oakland.edu},
\corauth[correspondingauthor]{Corresponding author.\quad\textsuperscript{\ref{equalcontribution}} Equal contribution.}
\author[Aero,equalcontribution]{Devansh R. Agrawal}\ead{devansh@umich.edu},
\author[Rob,Aero]{Dimitra Panagou}\ead{dpanagou@umich.edu}   
\address[Rob]{Department of Robotics, University of Michigan, Ann Arbor, MI 48109, USA}  
\address[Aero]{Department of Aerospace Engineering, University of Michigan, Ann Arbor, MI 48109, USA}

% \thanks[footnoteinfo]{Both Xinyi Wang and Devansh R. Agrawal contributed equally to this work. 
% E-mail addresses: xinywa@umich.edu (X. Wang), devansha@umich.edu (D.R. Agrawal), dpanagou@umich.edu (D. Panagou).}

                             % keyword list or with the 
                                          % help of the Automatica 
                                          % keyword wizard

\begin{abstract}
Stability analysis of the Kalman filter under randomly lost measurements has been widely studied. We revisit this problem in a general continuous-time framework, where both the measurement matrix and noise covariance evolve as random processes, capturing variability in sensing locations. Within this setting, we derive a closed-form upper bound on the expected estimation covariance for continuous-time Kalman filtering. We then apply this framework to spatiotemporal field estimation, where the field is modeled as a Gaussian process observed by randomly located, noisy sensors. Using clarity, introduced in our earlier work as a rescaled form of the differential entropy of a random variable, we establish a lower bound on the expected spatially averaged clarity and show that its steady-state value converges to a grid-independent limit as the grid is refined. This result exposes fundamental performance limits through a composite sensing parameter that jointly captures the effects of the number of sensors, noise level, and measurement frequency. Simulations show close agreement between the continuous-time analytical solution and the discrete-time covariance, with the agreement improving as the sampling interval decreases. The analytical solution also avoids the recursive computations required in the discrete-time formulation. Most importantly, the derived limits provide principled and efficient guidelines for sensor network design prior to deployment.
\begin{editone}\codelink{[Code]}\footnotemark[1]\end{editone}
\end{abstract}

\end{frontmatter}
\begingroup
\renewcommand{\thefootnote}{\fnsymbol{footnote}}
\setcounter{footnote}{1}
\refstepcounter{footnote}\label{equalcontribution}
\endgroup
\begingroup
\renewcommand{\thefootnote}{}
\footnotetext[0]{This work is being supported by the National Science Foundation (NSF) under Award Numbers 1942907 and 2223845. Xinyi Wang was a postdoctoral researcher at the University of Michigan when this work was conducted and is now with the Department of Mechanical Engineering, Oakland University, Rochester, MI 48309, USA.}
\endgroup
\footnotetext[1]{Code:~\codelink{https://github.com/Lawliet9666/multi-sensor-network-design}}
\setcounter{footnote}{1}

\section{Introduction}
Steady-state analysis of the estimation error reveals how uncertainty evolves and stabilizes over time \cite{anderson2005optimal}.
In traditional settings, the \ac{KF} provides the optimal linear minimum-variance estimator, with its steady-state covariance governed by a Riccati equation \cite{grewal2025kalman,lan2016rapidly,seo2025kalman}.
However, in modern applications such as estimating a spatiotemporal field using a multi-agent sensor network, the sensing process itself is inherently stochastic—sensor locations, measurement timing, and noise may all vary randomly \cite{marelli2019stability,dasgupta2025steady,sinopoli2004kalman}.
This randomness makes direct steady-state analysis of the estimation error intractable.
Consequently, it is more insightful to adopt a statistical viewpoint, by modeling sensing events as random processes and analyzing their collective impact on estimation performance.
Within this stochastic sensing framework, we address a fundamental sensor network design problem constrained by performance requirements:
\begin{editone}\emph{How should the number and quality of sensors be selected for a prescribed level of estimation accuracy prior to deployment?}\end{editone}
This motivates two key objectives: (1) to characterize the expected estimation performance analytically, and (2) to establish the fundamental limits of estimation accuracy under stochastic sensing.
In this work, we pursue these objectives through both theoretical analysis and application-driven validation.

Prior work has addressed this question in the discrete-time setting. In particular, Gupta \cite{gupta2006stochastic} analyzed the stochastic sensor selection problem and derived upper bounds on the expected error covariance of the \ac{KF} under randomly varying measurement availability.
While insightful, these bounds are recursive in nature and therefore computationally demanding. 
In this work, we approximate the measurement process using a continuous-time model rather than directly analyzing the discrete-time recursion. This formulation yields a differential inequality for the expected estimation covariance and a deterministic upper bound whose asymptotic behavior can be characterized in closed form.

The continuous-time approximation offers two main advantages.
First, the upper bound admits a closed-form steady-state expression, making the analysis computationally much more efficient than the recursive discrete-time formulation.
Second, in the context of multi-agent estimation of spatiotemporal \ac{GP} fields \cite{agrawal2024multi}, where agents collect noisy measurements at locations drawn from a spatial distribution, the \ac{GP} structure maps naturally to a linear state-space model, allowing the \ac{KF} to produce optimal estimates. 
To quantify estimation uncertainty within this framework, we adopt clarity, a rescaled form of the differential entropy of a random variable \cite{agrawal2023sensor}.
Leveraging the derived closed-form upper bound and its asymptotic behavior, the proposed approach reveals that the steady-state clarity lower bound converges to a grid-independent limit as the grid is refined.
This property enables the identification of the fundamental limits captured by this steady-state clarity lower bound under stochastic sensing. Consequently, the analysis provides a systematic tool for determining the minimal number of agents and the required sensing capabilities prior to deployment.
This directly addresses an optimal sensor resource design problem that is difficult to solve efficiently within the discrete-time formulation.

The main contributions are as follows:
\begin{itemize}
\item 
We develop an explicit closed-form upper bound on the expected estimation covariance for Kalman–Bucy filtering under randomized sensing.
Unlike the discrete-time formulation, this bound admits a closed-form steady-state expression, enabling efficient evaluation of estimation performance without recursive numerical computation.
\item 
We apply the proposed theoretical framework to spatiotemporal field estimation, where the field is modeled as a Gaussian process observed through randomly located, noisy sensors.
Within this context, clarity, derived from the differential entropy of a random variable, is employed to characterize estimation uncertainty.
Leveraging the established covariance bound, we derive a lower bound on the \begin{editone}expected spatially averaged clarity\end{editone}, revealing that the steady-state value of this lower bound is determined by sensing resources and captured by a single composite parameter that unifies the effects of the sensor number, measurement noise, and measurement rate.

\item 
Numerical simulations confirm that the proposed continuous-time bound becomes increasingly consistent with the discrete-time formulation as the measurement frequency increases. Its closed-form expression provides a practical sensor-number design criterion based on the continuous-time clarity lower bound and supports trade-off analysis among the sensor number, noise level, and sampling rate.

\end{itemize}

\section{Literature Review}

A substantial body of work has analyzed the covariance of the \ac{KF} estimation error, with particular emphasis on its steady-state behavior under various system conditions. For classical time-invariant systems, it is well established that under detectability and stabilizability, the covariance of estimation error converges to a steady state in both discrete- and continuous-time settings~\cite{lewis2017optimal}. 
Ni et al.~\cite{ni2016stability} further showed that, for systems with time-varying output matrices,
uniform complete observability alone ensures asymptotic stability of the continuous-time \ac{KF}.
However, the systems described above assume deterministic sensing. In contrast, modern sensing environments often deviate from this assumption due to inherent randomness: packet losses, stochastic sensor scheduling, or random sensor placement~\cite{sinopoli2004kalman,gupta2006stochastic,calle2021probabilistic}.

Sinopoli et al.~\cite{sinopoli2004kalman} pioneered the study of the \ac{KF} under random measurements by modeling packet losses as a Bernoulli process. They established statistical convergence properties of the expected covariance of estimation error and identified a critical observation-arrival rate, below which the covariance diverges and above which it remains bounded.

Gupta et al.~\cite{gupta2006stochastic} extended this  random measurement framework to a more general sensor scheduling problem. They proposed a stochastic sensor selection strategy that activates one sensor at each time step according to a prescribed probability distribution. Using the discrete-time Riccati recursion, they analyzed the evolution of the covariance of estimation error in expectation, deriving upper and lower bounds. Their results further demonstrated, through examples, that these bounds accurately approximate the actual steady-state covariance.

Subsequent research generalized these results to the setting of random measurement matrices. For example, Marelli et al.~\cite{marelli2019stability} considered the case where both the measurement matrix and the noise covariance are drawn randomly at each time from a known distribution, and established necessary and sufficient conditions for mean-square stability of the \ac{KF}. While closely related to our problem, their analysis is limited to stability criteria and does not provide bounds or steady-state characterizations of the error covariance.

Calle and Bopardikar~\cite{calle2021probabilistic} derived probabilistic semidefinite bounds on the covariance of estimation error for randomized sensor selection in Kalman Filtering. Their framework, however, models sensor selection as a one-time random sampling of a subset of sensors that remains fixed thereafter, in contrast to the per-step stochastic scheduling analyzed by Gupta et al.~\cite{gupta2006stochastic}.

Early approaches primarily relied on the discrete-time Riccati recursion to track the distribution or expectation of the covariance of estimation error. More recently, efforts have been made to analyze continuous-time process models with discrete-time measurement models~\cite{ahdab2025optimal,dasgupta2025steady,tanwani2020error}. For example, Ahdab et al.~\cite{ahdab2025optimal} modeled measurement arrivals as independent Poisson processes with sensor-specific rates and derived an upper bound on the mean posterior covariance of the continuous–discrete \ac{KF} along the mean auxiliary state.
Similarly, Dasgupta and Tanwani~\cite{dasgupta2025steady} studied continuous-time linear systems with Poisson-sampled observations. They showed that the Kalman–Bucy filter yields piecewise-deterministic covariance trajectories, and derived conditions on the mean observation rate that ensure bounded and convergent expected covariance of estimation error. Their results extend the discrete-time “dropout threshold” in~\cite{sinopoli2004kalman}.
However, their setting assumes a constant measurement matrix, with randomness arising from the irregular Poisson-distributed observation times rather than from the measurement matrix itself. In contrast, our focus is on a more general problem, akin to the framework of Marelli et al.~\cite{marelli2019stability}, where the measurement matrix itself is random.

Related work has also addressed continuous measurement models. For example, the work in~\cite{le2010scheduling} studies sensor scheduling for linear time-invariant systems. The authors derive a convex relaxation of the scheduling problem, yielding a performance lower bound. A lower bound reflects only the best-case performance and offers no guarantee on how the actual performance compares. In contrast, an upper bound constrains the worst-case covariance of the estimation error, providing a concrete performance guarantee.
Additionally, the work in~\cite{ha2019periodic} considers informative planning of sensing agents over an infinite time horizon for continuous-time linear systems. The covariance bound derived in this work tends to be conservative, as it omits the contribution of the measurement term.

Analysis of the covariance of the estimation error naturally informs the design of the sensing system. Once stability conditions or performance bounds have been established, a practical question arises: how should sensors be deployed or activated to guarantee a prescribed level of estimation accuracy? This leads to sensor scheduling formulations, in which the active subset is selected over time. Prior work has addressed these problems through design optimization grounded in covariance (or its bounds), deriving policies that guide per–time-step sensor activation~\cite{zhao2014optimal,ahdab2025optimal}.
Another typical problem is optimal sensor placement: given system parameters, one may wish to determine the minimum number of sensors or the required sensing quality needed to achieve a prescribed uncertainty threshold~\cite{tzoumas2016sensor}. Most prior work adopts covariance (or surrogates such as its log-determinant) as the performance metric and designs placement/scheduling policies accordingly.
However, covariance measures are often difficult to interpret or compare across different sensing configurations, since they are unit-dependent and different scalarizations (e.g., trace, log-determinant) can rank designs inconsistently. 
To evaluate the estimation quality, we employ clarity, as introduced in~\cite{agrawal2023sensor,agrawal2024multi}, which normalizes the differential entropy of a continuous random variable to the interval~$[0,1]$. 
In this paper, we show that within our closed-form upper-bound framework, the fundamental limits of estimation performance can be explicitly characterized, enabling direct quantification of the sensing resources required to achieve a prescribed uncertainty threshold.

\section{Problem Formulation}
We consider a continuous-time linear system evolving according to a \ac{SDE}:
\eqn{
\label{eq:continus_model}
\dot{\bm{x}}(t) = \bm{\bm{A}} \bm{x}(t) + \bm{w}(t), \quad \bm{w}(t) \sim \Ncal(\bm{0}, \bm{Q}_c),
}
where $\bm{x}(t) \in \R^n$ denotes the system state at time $t$, $\bm{\bm{A}} \in \R^{n \times n}$ is the system matrix, and $\bm{w}(t)$ is a zero-mean white Gaussian noise process with covariance $\bm{Q}_c\in \psd^n$.
The process state is observed by $N_r$ sensors whose spatial locations vary randomly over time within a compact spatial domain $D \subset \R^d$, where $d$ is the spatial dimension. 
Specifically, let
\begin{editone}
\eqn{
\label{eq:sensor_location}
\bm r(t):=
\big[\bm r_1(t)^\top,\ldots,\bm r_{N_r}(t)^\top\big]^\top
\in D^{N_r}\subset\R^{N_r d}
}
\end{editone}
\begin{editone}denote the stacked vector of sensor locations at time $t$, where $\bm r_j(t)\in D\subset\R^d$ is the location of sensor $j$.\end{editone}
\begin{editone}At each time $t$, the sensor locations are independently distributed according to the fixed spatial distribution $\mathcal{D}_r$:\end{editone}
$\begin{editone}\bm{r}_j(t) \overset{\mathrm{i.i.d.}}{\sim} \mathcal{D}_r,\, j=1,\ldots,N_r.\end{editone}
$
\begin{editone}The temporal dependence required for the continuous-time covariance bound is specified through the conditions on the induced information process $\bm{G}(t)$ in Theorem~\ref{thm:continuous_time_upper_bound}. The discrete-time sampling model used by the \ac{STGPKF} is specified separately in Section~5.\end{editone}
Each sensor measures a linear function of the process state corrupted by measurement noise:
\eqn{
\bm{y}(t) &= \bm{H}(t) \bm{x}(t) + \bm{v}(t), \quad \bm{v}(t) \sim \Ncal(0, \bm{V}_c(t))
}
where $\bm{y}(t) \in \R^{N_r}$ is the measurement vector, 
$\bm{H}(t) \in \R^{N_r \times n}$ is the time-varying observation matrix 
\begin{editone}with $\bm{H}(t)=\bm{H}(\bm r(t))$ and $\bm{V}_c(t)=\bm{V}_c(\bm r(t))$ both determined by the instantaneous sensor-location vector $\bm r(t)$\end{editone}, and
$\bm{v}(t)$ is a zero-mean white Gaussian measurement-noise process with intensity matrix $\bm{V}_c(t)$.
The measurement noises are assumed independent of the process noise $\bm{w}(t)$.
\begin{editone}
We assume that the measurements collected by all $N_r$ sensors are available
through a common observation history, based on which a common state estimate
$\hat{\bm{x}}(t)$ is formed. Under this shared-information assumption, we study
how the estimation performance depends on $N_r$ and determine the number of
sensors required to achieve a prescribed target clarity.
\end{editone}

% Since all sensors share their measurements, each has access to the same global observation history and maintains a common estimate of the system state, denoted by $\hat{\bm{x}}(t)$.

\begin{editone}Let $\mathcal{P}_t := \sigma\{\bm r(s):0\leq s\leq t\}$ denote the information generated by the sensing history up to time \(t\).\end{editone}
Given the measurement model above, the optimal estimate of $\bm{x}(t)$ is provided by the Kalman–Bucy filter, whose \begin{editone}conditional covariance matrix\end{editone}
$
\begin{editone}\bm{\Sigma}(t) := \E\!\left[(\bm{x}(t) - \hat{\bm{x}}(t))(\bm{x}(t) - \hat{\bm{x}}(t))^{\top}\mid\mathcal{P}_t\right]\end{editone}
$
evolves according to the Riccati differential equation
\eqnOne{
\label{eq:continuous_riccati}
\dot{\bm{\Sigma}}(t) 
&= \bm{A}\bm{\Sigma}(t) + \bm{\Sigma}(t)\bm{A}^{\top} + \bm{Q}_c \\
&- \bm{\Sigma}(t)\bm{H}(t)^{\top}\bm{V}_c(t)^{-1}\bm{H}(t)\bm{\Sigma}(t),
}
where $\bm{\Sigma}(0) = \bm{\Sigma}_0 \in \psd^{n}$ is the initial covariance matrix.

For notational convenience, we introduce the Riccati operator 
parameterized by $\bm{G} \in \psd^n$ as
\eqnOne{
\label{eq:riccati_op}
\Ric_{G}(\bm{\Sigma}) 
:= \bm{\bm{A}} \bm{\Sigma} + \bm{\Sigma} \bm{\bm{A}}^\top + \bm{Q}_c - \bm{\Sigma} \bm{G} \bm{\Sigma},
}
where $\bm{\Sigma} \in \pd^n$.
With this notation, the Kalman–Bucy Riccati equation reads
\eqnOne{
\label{eq:continuous_information_matrix}
\dot{\bm{\Sigma}}(t) = \Ric_{\bm{G}(t)}(\bm{\Sigma}(t)),
\quad 
\bm{G}(t) := \bm{H}(t)^\top  \bm{V}_c(t)^{-1} \bm{H}(t).
}
Because the sensor-location vector $\bm r(t)$ evolves randomly, both $\bm{H}(t)$ and $\bm{V}_c(t)$ are random matrix-valued processes, rendering $\bm{\Sigma}(t)$ itself a random matrix process. Our objective is to characterize the expected covariance $\bar{\bm{\Sigma}}(t) := \E_{\mathcal{P}_t}[\bm{\Sigma}(t)]$ and to derive tractable upper bounds on $\bar{\bm{\Sigma}}(t)$ that capture how stochastic sensing influences estimation performance.
For brevity, we will write $\E[\cdot]$ to denote expectation with respect to
the sensing history $\mathcal{P}_t$.

\section{Continuous-Time Upper Bound and Analytical Analysis}

\subsection{Continuous-Time Upper Bound of Expected Covariance}
To establish the continuous-time upper bound, we first derive a set of structural properties of the Riccati operator that enable us to bound $\bar{\bm{\Sigma}} (t)$.
\begin{lem}\textbf{[Concavity]}
\label{lem:concavity}
The operator $\Ric_{G}$ is concave, that is,  for any $\bm{\Sigma}_1, \bm{\Sigma}_2 \in \pd^n$ and $\beta \in [0, 1]$, 
\eqn{
\Ric_{G}(\beta \bm{\Sigma}_1 \!+\!(1- \!\beta)\bm{\Sigma}_2) \succeq \beta \Ric_{G}(\bm{\Sigma}_1) \!+\! (1 -\! \beta) \Ric_{G}(\bm{\Sigma}_2)
}
\end{lem}
\begin{pf}
    See Appendix~\ref{appendix:concavity}.
\end{pf}

\begin{lem}
\label{lem:diff_of_riccati}
    For any $\bm{\Sigma}_1, \bm{\Sigma}_2 \in \psd^n$,  let $\bm{K} = \bm{\Sigma}_1 - \bm{\Sigma}_2$. Then,
    $
    \Ric_{G}(\bm{\Sigma}_1) - \Ric_{G}(\bm{\Sigma}_2) = \tilde{\bm{A}} \bm{K}  +  \bm{K} \tilde{\bm{A}}^\top
    $
    where $\tilde{\bm{A}} = \bm{A} - \frac{1}{2}  (\bm{\Sigma}_1 + \bm{\Sigma}_2) \bm{G}$.
\end{lem}
\begin{pf}
    See Appendix~\ref{appendix:diff_of_riccati}.
\end{pf}

\begin{lem}
\label{lem:lyapunov}
    Let $\tilde{\bm{A}} : [t_0, \infty) \to \R^{ n \times n}$. Suppose $\bm{K}: [t_0, \infty) \to \R^{n \times n}$ satisfies  
    \eqn{
    \dot{\bm{K}}(t) \succeq  \tilde{\bm{A}}(t) \bm{K}(t) + \bm{K}(t) \tilde{\bm{A}}(t)^\top,  \quad \forall t \geq t_0,
    }
    with  initial condition $\bm{K}(t_0)$.
    Then, 
    $
    \bm{K}(t_0) \succeq 0 \implies \bm{K}(t) \succeq 0, \,\forall t \geq t_0.
    $
\end{lem}
\begin{pf}
    See Appendix~\ref{appendix:lyapunov}.
\end{pf}

\begin{thm}\textbf{[Continuous-Time Upper Bound]}
\label{thm:continuous_time_upper_bound}
Let the averaged information matrix be
\eqn{
\label{eq:g_bar_expected}
% \bar{\bm{G}} := \E_{\bm r(t)}[\bm{H}(t)^\top \bm{V}_c(t)^{-1}\bm{H}(t)].
\bar{\bm{G}} := \E[\bm{H}(t)^\top \bm{V}_c(t)^{-1}\bm{H}(t)].
}
\begin{editone}Assume that $\bar{\bm{G}}$ is time invariant and that
$\E[\bm{G}(t)\mid\bm{\Sigma}(t)]=\bar{\bm{G}}$ for almost every $t\geq t_0$.\end{editone}
% \edit{Assume also that
% $\frac{d}{dt}\E[\bm{\Sigma}(t)]=\E[\dot{\bm{\Sigma}}(t)]$
% for almost every $t\geq t_0$.}
% \edit{One exact case satisfying $\E[\bm{G}(t)\mid\bm{\Sigma}(t)]=\bar{\bm{G}}$ is $\bm{G}(t)=\bar{\bm{G}}$ for almost every $t$.}
Define $\bm{\Delta}:[t_0,\infty)\to\pd^n$ as the solution of the Riccati differential equation
\eqn{
\label{eq:Delta}
\dot{\bm{\Delta}} = \bm{A}\bm{\Delta}+\bm{\Delta} \bm{A}^\top+\bm{Q}_c-\bm{\Delta}\bar{\bm{G}}\bm{\Delta}, 
\quad \bm{\Delta}(t_0)=\bar{\bm{\Sigma}}(t_0).
}
Then, for all $t\geq t_0$, $\bar{\bm{\Sigma}} (t) \preceq \bm{\Delta}(t).$
\end{thm}
\begin{pf}
Differentiating $\bar{\bm{\Sigma}}(t)=\E[\bm{\Sigma}(t)]$ gives
    \seqn{
    % \frac{d}{dt} \E[\bm{\Sigma}(t)] 
     \dot {\bar{\bm{\Sigma}}} 
    & = \E[ \dot{\bm{\Sigma}}(t)] \label{eq:1}\\
    &= \E[ \bm{A} \bm{\Sigma}(t) + \bm{\Sigma}(t) \bm{A}^\top + \bm{Q}_c - \bm{\Sigma}(t) \bm{G}(t) \bm{\Sigma}(t)] \label{eq:2}\\
    &= \E[ \bm{A} \bm{\Sigma}(t) + \bm{\Sigma}(t) \bm{A}^\top + \bm{Q}_c- \bm{\Sigma}(t) \bar{\bm{G}} \bm{\Sigma}(t)] \label{eq:3}\\
    &= \E [ \Ric_{\bar{G}} ( \bm{\Sigma}(t)) ] \label{eq:4}\\
    &\preceq \Ric_{\bar{G}}(\E[\bm{\Sigma}(t)])\label{eq:5}
    }
    where Eq.~\eqref{eq:1} follows from Fubini's theorem, since the Riccati drift is integrable on every finite time interval; Eq.~\eqref{eq:2} follows by substituting the expression for $\dot{\bm{\Sigma}}$;
    \begin{editone}Eq.~\eqref{eq:3} follows from the tower property and the assumption
    $\E[\bm{G}(t)\mid\bm{\Sigma}(t)]=\bar{\bm{G}}$, since
    $\E[\bm{\Sigma}(t)\bm{G}(t)\bm{\Sigma}(t)]
    =\E[\bm{\Sigma}(t)\E[\bm{G}(t)\mid\bm{\Sigma}(t)]\bm{\Sigma}(t)]$
    and $\E[\bm{G}(t)\mid\bm{\Sigma}(t)]=\bar{\bm{G}}$;\end{editone}
    Eq.~\eqref{eq:4} invokes the definition of the Riccati operator in Eq.~\eqref{eq:riccati_op}, 
    and Eq.~\eqref{eq:5} follows from Jensen’s inequality together with the concavity of $\Ric_{G}$ established in Lemma~\ref{lem:concavity}. 
    
    Define $\bm{K}(t) = \bm{\Delta}(t) - \bar{\bm{\Sigma}}(t)$. Then,
    \eqnOne{
    \dot{\bm{K}}(t) &= \dot{\bm{\Delta}}(t) - \dot{\bar{\bm{\Sigma}}}(t)\succeq \Ric_{\bar{G}}(\bm{\Delta}(t)) - \Ric_{\bar{G}}(\bar{\bm{\Sigma}} (t))\\
    &= \tilde{\bm{A}}(t) \bm{K}(t)  +   \bm{K}(t) \tilde{\bm{A}}(t)^\top
    }
    where $\tilde{\bm{A}}(t) = \bm{A} - \frac{1}{2} ( \bm{\Delta}(t) + \bar{\bm{\Sigma}}(t)) \bar{\bm{G}}$, based on the proof in Lemma~\ref{lem:diff_of_riccati}. By Lemma~\ref{lem:lyapunov}, this means that $\bm{K}(t) \succeq 0$ for all $t \geq t_0$.
\end{pf}

\begin{cor}
Let $\Lcal: \psd^n \to \R$ be \begin{editone}order-preserving with respect to the Loewner order\end{editone}. Then
\eqn{
\Lcal(\bar{\bm{\Sigma}}(t)) \leq \Lcal( \bm{\Delta}(t)), \quad \forall t \geq t_0. 
}
\end{cor}

\begin{rem}
Theorem~\ref{thm:continuous_time_upper_bound} provides, to the best of our knowledge,  a continuous-time upper bound on the expected estimation covariance $\bar{\bm{\Sigma}}(t)$ under the stated conditions. For each $t\geq t_0$, the expected covariance is bounded by the deterministic solution $\bm{\Delta}(t)$, which evolves according to the Riccati differential equation in Eq.~\eqref{eq:Delta}.
\end{rem}

\subsection{Closed-form Solution of Upper Bound}

To gain further insight into the structure of this bound, we next consider a canonical case in which the system dynamics and process excitation are isotropic, i.e., $\bm{A} = a\bm{I}$ with $a < 0$ and $\bm{Q}_c = q_c\bm{I}$. 
Under this assumption, the upper-bound Riccati dynamics decouple in the eigenbasis of the averaged information matrix $\bar{\bm{G}}$, reducing to a set of scalar Riccati equations, one along each eigendirection of $\bar{\bm{G}}$ as defined in Eq.~\eqref{eq:g_bar_expected}.
This simplification enables a closed-form characterization of the steady-state solution of the upper-bound Riccati equation, as stated in the following theorem.

\begin{thm}\textbf{[Closed-Form of Upper Bound]}
\label{thm:analystical_sol_continuous_riccati}
Consider the continuous-time differential equation governing the upper bound $\bm{\Delta}$ as:
\eqnOne{
&\dot{\bm{\Delta}} \;=\; \bm{A}\bm{\Delta}+\bm{\Delta} \bm{A}^\top + \bm{Q}_c - \bm{\Delta} \bar{\bm{G}} \bm{\Delta},\\
&  \bm{A}=a\bm{I}~(a<0),\; \bm{Q}_c=q_c \bm{I}~(q_c>0),\; \bar{\bm{G}}\in\pd^n.
}
Then $\bm{\Delta}(t)$ converges to the unique stabilizing solution $\bm{\Delta}_\infty$ of the \ac{CARE}
\eqn{
\label{eq:care}
0 \;=\; \bm{A} \bm{\Delta}_\infty + \bm{\Delta}_\infty \bm{A}^\top + \bm{Q}_c \bm{I} - \bm{\Delta}_\infty \bar{\bm{G}} \bm{\Delta}_\infty,
}
given in closed-form by
\eqn{
\bm{\Delta}_\infty = - q_c \left( a\bm{I}-\sqrt{a^{2}\bm{I}+q_c\bar{\bm{G}}} \right)^{-1}.
}
Equivalently, if $\{\lambda_i\}_{i=1}^n$ are the eigenvalues of $\bar{\bm{G}}$, then the eigenvalues
of $\bm{\Delta}_\infty$ are
\eqn{
\label{eq:gamma}
\gamma_i =\frac{-q_c}{a-\sqrt{a^2+q_c\lambda_i}},
\quad i=1,\dots,n.
}
\end{thm}
\begin{pf}
Since $\bar{\bm{G}} \in \S_{++}^n$, it admits the eigen-decomposition $\bar{\bm{G}} = \bm{U} \bm{\Lambda}_G \bm{U}^\top$, where $\bm{U}$ is orthogonal and $\bm{\Lambda}_G$ is diagonal. 
Under the isotropic assumption $\bm{A} = a\bm{I}$ and $\bm{Q}_c = q_c \bm{I}$, Eq.~\eqref{eq:care} implies that $\bar{\bm{G}}$ and $\bm{\Delta}_\infty$ commute, i.e., $\bar{\bm{G}}\bm{\Delta}_\infty = \bm{\Delta}_\infty\bar{\bm{G}}$. 
By Theorem~4.5.15 (a) in \cite{horn2012matrix}, any two commuting symmetric matrices can be simultaneously diagonalized by the same orthogonal matrix. 
Hence, 
% $\bm{\Delta}_\infty = \bm{U} \bm{\Lambda}_{\Delta} \bm{U}^\top$, where $\bm{\Lambda}_{\Delta}$ is diagonal. 
applying the same factorization to $\bm{\Delta}_\infty$, we have $\bm{\Delta}_\infty = \bm{U} \bm{\Lambda}_{\Delta} \bm{U}^\top$, where $\bm{\Lambda}_{\Delta}$ is diagonal.
    % by Lemma~\ref{lem:gamma_diagonal}. % (\note{todo: need to prove})
    Therefore, Eq.~\eqref{eq:care} becomes
    \eqnN{
    0 &= \bm{A} \bm{\Delta}_\infty + \bm{\Delta}_\infty \bm{A}^\top + \bm{Q}_c - \bm{\Delta}_\infty \bar{\bm{G}} \bm{\Delta}_\infty\\
    &= 2a \bm{U} \bm{\Lambda}_{\Delta} \bm{U}^\top + q_c \bm{I}
    - \bm{U}\bm{\Lambda}_{\Delta}\bm{\Lambda}_G\bm{\Lambda}_{\Delta}\bm{U}^\top\\
    &= q_c \bm{I} + \bm{U}( 2 a \bm{\Lambda}_{\Delta} - \bm{\Lambda}_{\Delta} \bm{\Lambda}_G \bm{\Lambda}_{\Delta}) \bm{U}^\top\\
    &= \bm{U}(q_c \bm{I} + 2 a \bm{\Lambda}_{\Delta} - \bm{\Lambda}_{\Delta} \bm{\Lambda}_G \bm{\Lambda}_{\Delta}) \bm{U}^\top
    }
    where the term inside the parentheses is a diagonal matrix. For the equation to hold, each diagonal entry must be zero. Let $\gamma_i = [\bm{\Lambda}_{\Delta}]_{ii}$ and $\lambda_{i} = [\bm{\Lambda}_G]_{ii}$. Then,
    $q_c + 2 a \gamma_i - \lambda_i \gamma_i^2 = 0,
    $
    which has the solutions
    $
    \gamma_i = \frac{a \pm \sqrt{ a^2 + q_c \lambda_i}}{\lambda_i}.
    $
    Since we seek the stabilizing solution, we keep only the positive solutions, i.e.,
    \eqn{
    \gamma_i = \frac{a+\sqrt{a^2+q_c\lambda_i}}{\lambda_i} =\frac{-q_c}{a-\sqrt{a^2+q_c\lambda_i}}
\quad i=1,\dots,n.
    }
    Thus, we have derived the eigenvalues of $\bm{\Delta}_\infty$. 
   Using these eigenvalues, we have 
    \eqnOne{
    \bm{\Delta}_\infty &= \bm{U} \bm{\Lambda}_\Delta  \bm{U}^\top = \bm{U} \big( \operatorname{diag}( \gamma_1, ..., \gamma_n) \big) \bm{U}^\top \\
    &= \bm{U} \big( a \bm{\Lambda}_G^{-1} + \bm{\Lambda}_G^{-1/2} (a^2 \bm{I} + q_c \bm{\Lambda}_G)^{1/2} \bm{\Lambda}_G^{-1/2} \big)\bm{U}^\top.
    }
    This can be expressed succinctly as
    \eqn{
    \bm{\Delta}_\infty = \bar{\bm{G}}^{-1/2}  \left( a \bm{I} + ( a^2 \bm{I} + q_c \bar{\bm{G}})^{1/2} \right) \bar{\bm{G}}^{-1/2}
    }
    Since $\bar{\bm{G}} \in \pd^n$, $(a^{2}\bm{I}+q_c\bar{\bm{G}})^{1/2}$ is a closed-form matrix function of $\bar{\bm{G}}$ and thus commutes with $\bar{\bm{G}}^{\pm1/2}$. Hence,
    \eqnOne{
    \bm{\Delta}_\infty
    &= a\,\bar{\bm{G}}^{-1/2}\bm{I}\bar{\bm{G}}^{-1/2}
      + (a^{2}\bm{I}+q_c\bar{\bm{G}})^{1/2}\bar{\bm{G}}^{-1/2}\bar{\bm{G}}^{-1/2}\\
    &= \left( a\bm{I}
      + (a^{2}\bm{I}+q_c\bar{\bm{G}})^{1/2} \right) \bar{\bm{G}}^{-1}\\
    &= -q_c\left( a\bm{I}-\sqrt{a^{2}\bm{I}+q_c\bar{\bm{G}}} \right)^{-1}.
    }
\end{pf}

\begin{rem}
Theorem~\ref{thm:analystical_sol_continuous_riccati} provides an explicit closed-form expression for the steady-state upper bound on the expected covariance of the estimation error. 
The solution reveals how the eigenvalues of the averaged information matrix $\bar{\bm{G}}$ directly determine the achievable estimation accuracy, offering a tractable and interpretable characterization of continuous-time estimation performance.
\end{rem}

Based on the above analysis, understanding what factors determine~$\bar{\bm{G}}$ is key to explaining how sensing resources influence estimation performance.
This connection motivates the following section, in which we derive a closed-form characterization of~$\bar{\bm{G}}$ under a spatiotemporal sensing environment.

\section{Application to Spatiotemporal Estimation}
The continuous-time upper-bound framework developed in the previous section provides a general characterization of estimation uncertainty under stochastic sensing. 
We now consider the problem of spatiotemporal field estimation, where the process of interest is modeled as a \ac{GP} evolving continuously in both space and time. 
To estimate the spatiotemporal field from noisy measurements collected at randomized sensor locations, we employ the~\ac{STGPKF} proposed in~\cite{agrawal2024multi}.
\begin{editone}In this section, we use the continuous-time upper-bound results derived earlier as an analytically tractable surrogate for the recursive covariance analysis of the \ac{STGPKF}.\end{editone}
% Simulation results are finally provided to illustrate the theoretical predictions.
% \subsection{Problem Description}

An overview of main notations used in this section is provided in Table~\ref{tab:symbols}.

\begin{table}[t]
\centering
\caption{Summary of covariance- and clarity-related symbols.}
\label{tab:symbols}
\renewcommand{\arraystretch}{1.0}
\setlength{\tabcolsep}{4pt}
\begin{tabularx}{\columnwidth}{@{}L{0.26\columnwidth} X@{}}
\toprule
\textbf{Symbol} & \textbf{Definition} \\
\midrule
\multicolumn{2}{l}{\textit{Covariance-related symbols}} \\
\midrule
$\bm{\Sigma}$ & Covariance of the latent-state estimation error $\bm{s} - \hat{\bm{s}}$. \\
$\bar{\bm{\Sigma}} =\E[\bm{\Sigma}]$ & Expected covariance of the latent-state estimation error under stochastic sensing. \\
$\bm{\Delta}$ & Upper bound of $\bar{\bm{\Sigma}}$. \\
$\bm{\Pi}$ & Covariance of the field-state estimation error $\bm{f} - \hat{\bm{f}}$. \\
$\bar{\bm{\Pi}} = \E[\bm{\Pi}]$ & Expected covariance of the field-state estimation error under stochastic sensing.  \\
$\bm{\Delta}^\Pi$ & Upper bound of $\bar{\bm{\Pi}}$. \\
$\bm{\Delta}_\infty$ & Steady-state solution of the upper-bound Riccati equation. \\
\midrule
\multicolumn{2}{l}{\textit{Clarity-related symbols}} \\
\midrule
$q_i$ & Clarity at the $i$-th spatial location. \\
$\bar{q}$ & Spatially averaged clarity for a given sensor realization. \\
\begin{editone}$\E[\bar q]$\end{editone} & \begin{editone}Expected spatially averaged clarity defined in Eq.~\eqref{eq:expected_clarity}.\end{editone} \\
\begin{editone}$\bar{q}_{\Delta^\Pi}$\end{editone} & \begin{editone}Lower bound on expected spatially averaged clarity defined in Eq.~\eqref{eq:q_delta}.\end{editone} \\
\begin{editone}$\bar{q}_{\Delta^{\bm{\Pi}}_\infty}$\end{editone} & \begin{editone}Steady-state lower bound on expected spatially averaged clarity defined in Eq.~\eqref{eq:clarity_steady_state}.\end{editone} \\
\bottomrule
\end{tabularx}
\end{table}

\subsection{Spatiotemporal Environment}
\label{sec:spatio_temporal_env}
We consider a spatiotemporal field $f : \R \times D \to \R$, modeled as a \ac{GP} with a separable kernel
\eqn{
\label{eq:f}
    f(t,p) \sim \mathcal{GP}\big(0,\, k_T(t,t')\,k_S(\bm{p},\bm{p}')\big), \, \bm{p} \in D \subset \R^d,
}
where $k_T:\R\times\R\to\R$ and $k_S:D\times D\to\R$ are continuous, symmetric, positive definite kernels capturing temporal and spatial correlations, respectively. 
For example, we may use the Matérn-$\tfrac{1}{2}$ kernel in both time and space:
\eqn{
k_T(t,t') &= \sigma_t^2 \exp\!\left(-\tfrac{|t-t'|}{l_t}\right),  \\
k_S(\bm{p},\bm{p}') &= \sigma_s^2 \exp\!\left(-\tfrac{\|\bm{p}-\bm{p}'\|}{l_s}\right), \label{eq:ks}
}
with variances $\sigma_t^2,\sigma_s^2$ and correlation lengths $l_t,l_s$.

Agents act as sensors that probe this \ac{GP} field. 
At a fixed sampling period $\Delta t >0$, measurements are collected by $N_r$ agents whose sampling locations vary randomly. 
\begin{editone}Specifically, at the $k$-th timestep $t_k=k\Delta t$, define
$\bm r_k:=\bm r(t_k)=
[\bm r_{1,k}^\top,\ldots,\bm r_{N_r,k}^\top]^\top$,
where $\bm r_{j,k}:=\bm r_j(t_k)$ and
$\bm r_{j,k}\overset{\mathrm{i.i.d.}}{\sim}\mathcal D_r$,
$j=1,\ldots,N_r$. The sampled location vectors $\bm r_k$ are independent across sampling times.\end{editone}
% Here $r_j(t_k) \in D \subset \R^d$ denotes the location of the $j$-th agent at time $t_k$.
\begin{editone}At each location, the $j$-th agent acquires a noisy measurement at time $t_k$:\end{editone}
\begin{editone}
\eqn{
\label{eq:sensing_model}
    y_{j}(t_k) \;=\; f(t_k, \bm r_{j,k}) + \eta_{j,k},
    \, \eta_{j,k} \sim \mathcal{N}\!\left(0, \sigma_m^2\right).
}
\end{editone}
% where 
\begin{editone}Here $y_j(t_k) \in \R$ is the scalar measurement reported by agent $j$, and $\eta_{j,k}$ is zero-mean Gaussian noise. The parameter
$
\sigma_m^2:=\frac{\sigma_c^2}{\Delta t},
\, \sigma_c^2>0,
$
% $\sigma_m^2$ 
is the noise variance of one discrete measurement collected with sampling period $\Delta t$, while $\sigma_c^2$ is the fixed continuous-time measurement-noise intensity.
The noise variables $\eta_{j,k}$ are mutually independent across agents and independent of the underlying GP process.
\end{editone}

\subsection{Spatiotemporal Gaussian Process Kalman Filter}

To estimate the spatiotemporal field $f(t,\bm{p})$ from the randomized sensor measurements defined in Eq.~\eqref{eq:sensing_model}, we begin by introducing a finite-dimensional 
approximation of the spatial domain. 
For grid spacing $\delta>0$, define the uniform grid
\eqn{
\label{eq:x_grid}
P_g^{\delta} = \{\bm{p}^{\delta}_1,\dots,\bm{p}^{\delta}_{N_g^{\delta}}\} \subset D,
}
where $\bm{p}^{\delta}_i \in D \subset \R^d$ denotes the location of the $i$th grid point. The total number of grid points is given by~$N_g^{\delta} := \frac{|D|}{\delta^d}$, where $|D|>0$ is the Lebesgue measure of $D$.

Following previous work in~\cite{agrawal2024multi}, the temporal kernel $k_T$ admits an equivalent finite-dimensional \ac{SDE} representation.
At each $\bm{p}^{\delta}_i$, let $\bm{s}_i(t)\in\mathbb{R}^{n_k}$ denote the latent temporal state induced by the temporal kernel, where $n_k$ is the state dimension. Stacking yields $\bm{s}(t) := [\bm{s}_1(t)^\top,\ldots,\bm{s}_{N_g^{\delta}}(t)^\top]^\top \in \R^{N_g^{\delta} n_k}$.
The latent state evolves according to
\eqnOne{
\label{eq:process_model_continuous}
d\bm{s}(t) &= \big(\bm{I}_{N_g^{\delta}} \otimes \bm{A}_0\big)\,\bm{s}(t)\,dt
               + \big(\bm{I}_{N_g^{\delta}} \otimes \bm{B}_0\big)\,d\bm{W}(t) \\
               &= \bm{A}\,\bm{s}(t)\,dt + \bm{B}\,d\bm{W}(t),
}
where $\bm{W}(t)=[W_1(t),\ldots,W_{N_g^{\delta}}(t)]^\top$ collects independent standard Wiener processes,
$\bm{A}_0\in\R^{n_k\times n_k}$ and $\bm{B}_0\in\R^{n_k\times 1}$ are the state and diffusion matrices for a single grid point determined by the chosen temporal kernel, and $\bm{A}:=\bm{I}_{N_g^{\delta}} \otimes \bm{A}_0$, $\bm{B}:=\bm{I}_{N_g^{\delta}} \otimes \bm{B}_0$.
The continuous-time process model in Eq.~\eqref{eq:process_model_continuous} is equivalent to Eq.~\eqref{eq:continus_model} with the process noise covariance defined as $\bm{Q}_c = \bm{B}\bm{B}^\top$ according to \cite{sarkka2013gaussian}. 
For the Matérn-$\tfrac{1}{2}$ kernel ($n_k=1$), $\bm{A}_0$ and $\bm{B}_0$ reduce to scalars, yielding the isotropic process dynamics considered in the continuous-time analysis.

The spatial correlations among grid points are characterized by the kernel matrix
\eqn{
\label{eq:kgg_delta}
\bm{K}_{gg}^{\delta} \in \R^{N_g^{\delta}\times N_g^{\delta}},
\, \big[\bm{K}_{gg}^{\delta}\big]_{ij} = k_S \big(\bm{p}^{\delta}_i,\bm{p}^{\delta}_j\big).
}
% \textit{Notation:}  
For brevity, when the grid spacing $\delta$ is fixed we drop the superscript and write
$\bm{K}_{gg} := \bm{K}_{gg}^{\delta}, \, N_g := N_g^{\delta}, \, \bm{p}_i := \bm{p}_i^{\delta}$.

The field values at the grid points are obtained from the latent state via
\eqnOne{
\label{eq:field_state}
\bm{f}(t) =\sqrt{\bm{K}_{gg}}(\bm{I}_{N_g}\otimes \bm{C}_0)\bm{s}(t)
=\sqrt{\bm{K}_{gg}}\bm{C}\bm{s}(t)
}
where $\bm{C}_0 \in \mathbb{R}^{1 \times n_k}$ is the local observation matrix that maps 
the latent temporal state $\bm{s}_i(t) \in \mathbb{R}^{n_k}$ 
at grid point $i$ to the scalar field value $f_i(t) \in \mathbb{R}$, and $\bm{C} := \bm{I}_{N_g} \otimes \bm{C}_0$.

While the process model captures \ac{GP} dynamics on a discretized grid, agents can acquire noisy measurements at random off-grid locations.
To relate the latent grid-based state $\bm{s}(t)$ to these physical observations, we introduce the following spatial kernel blocks:
\seqn{
    \bm{K}_{gr}(\bm r_k) &= \big[k_S(\bm p_i,\, \bm r_{j,k})\big]_{i,j}, \\
    \bm{K}_{rg}(\bm r_k) &= \bm{K}_{gr}^\top(\bm r_k), \\
    \bm{K}_{rr}(\bm r_k) &= \big[k_S(\bm r_{i,k},\, \bm r_{j,k})\big]_{i,j}.
}
Conditioning the joint Gaussian over grid and agent locations yields the discrete-time linear measurement model
\eqnOne{
\label{eq:measurement_model}
&\bm{y}(t_k) 
= \bm{H}(\bm r_k)\,\bm{s}(t_k) + \bm{v}(t_k), \\
&\bm{v}(t_k) \sim \mathcal{N}\!\big(\bm{0},\bm{V}(\bm r_k)\big),
}
with
$
\bm{H}(\bm r_k) = \bm{K}_{rg}(\bm r_k)\,\bm{K}_{gg}^{-1}\,\sqrt{\bm{K}_{gg}} \bm{C}, \quad
\bm{V}(\bm r_k) = \sigma_m^2 \bm{I}_{N_r} + \bm{K}_{rr}(\bm r_k) 
- \bm{K}_{rg}(\bm r_k) \bm{K}_{gg}^{-1} \bm{K}_{gr}(\bm r_k).
$
% \edit{Here $(\sigma_c^2/\Delta t)\bm{I}_{N_r}$ is the sensor-noise covariance specified in Eq.~\eqref{eq:sensing_model}. Because the GP kernel in Eq.~\eqref{eq:f} is separable, the second term is the same-time spatial conditional covariance of the field values at the agent locations given those at the grid points; for the kernel in Eq.~\eqref{eq:ks}, $k_T(t_k,t_k)=\sigma_t^2$. In the discrete-time STGPKF, these conditional residuals are modeled as independent across sampling times and independent of the state-process noise. Under this finite-grid measurement approximation, $\bm{V}(\bm{r}(t_k))$ is the effective discrete-time measurement-noise covariance.}
Here $\bm H(\bm r_k)$ is the sensing matrix and $\bm V(\bm r_k)$ is the measurement noise covariance,
both determined by the sensor-location vector $\bm r_k$ sampled at $t_k$ and the measurement noise variance $\sigma_m^2$.
For compactness, we adopt the shorthand $\bm H_k:=\bm H(\bm r_k)$, $\bm V_k:=\bm V(\bm r_k)$, $\bm s_k:=\bm s(t_k)$, and $\bm y_k:=\bm y(t_k)$.

For the continuous-time process model and discrete measurement model defined in~\eqref{eq:process_model_continuous} and~\eqref{eq:measurement_model}, we follow the approach in~\cite{agrawal2024multi}, which discretizes the continuous-time dynamics into an equivalent discrete-time state–space representation (see, e.g.,~\cite{wahlstrom2014discretizing}):
\eqnOne{
\label{eq:process_model_discrete}
    \bm{s}_{k+1} = \Phi\,\bm{s}_{k} + \bm{w}_k, \, \bm{w}_k \sim \mathcal{N}(\bm{0},\bm{Q}_d),
}
where the transition and process-noise matrices are defined as
$\bm{\Phi} := \bm{I}_{N_g^{\delta}} \otimes \Phi_0,\, \bm{Q}_d := \bm{I}_{N_g^{\delta}} \otimes \bm{Q}_{d_0} $
with 
\eqnOne{
\Phi_0 = e^{\bm{A}_0 \Delta t}, \quad
\bm{Q}_{d_0} = \int_0^{\Delta t} e^{\bm{A}_0\tau}\bm{B}_0\bm{B}_0^{\top}e^{\bm{A}_0^{\top}\tau} d\tau 
}
% \edit{Under the stated discrete-time noise model, Eqs.~\eqref{eq:measurement_model} and~\eqref{eq:process_model_discrete} define a standard linear Gaussian state-space model. Conditioned on the time-varying sensing matrices, the \ac{KF} provides its minimum mean-square error state estimate.}
As shown in~\cite{agrawal2024multi}, the resulting discrete-time \ac{KF} provides the optimal state estimate, and Eqs.~\eqref{eq:process_model_discrete} and~\eqref{eq:measurement_model} together define a standard linear discrete-time \ac{KF}.
Let $(\hat{\bm{s}}_{k|k},\, \bm{\Sigma}_{k|k})$ denote the posterior mean and covariance 
of the latent state at step $k$.  
Conditioned on the time-varying sensing pair $(\bm{H}_k,\bm{V}_k)$, the \ac{KF} update is:

\textit{Update step:}
\eqnOne{
\bm{S}_k &= \bm{H}_k\, \bm{\Sigma}_{k|k-1}\, \bm{H}_k^\top + \bm{V}_k, \\
\bm{K}_k &= \bm{\Sigma}_{k|k-1}\, \bm{H}_k^\top\, \bm{S}_k^{-1}, \\
\hat{\bm{s}}_{k|k} &= \hat{\bm{s}}_{k|k-1} + \bm{K}_k \big( \bm{y}_k - \bm{H}_k \hat{\bm{s}}_{k|k-1} \big), \\
\label{eq:kf_update}
\bm{\Sigma}_{k|k} &= \big(\bm{I} - \bm{K}_k \bm{H}_k\big)\, \bm{\Sigma}_{k|k-1}.
}
\textit{Prediction step:}
\eqnOne{
\hat{\bm{s}}_{k+1|k} &= \bm{\Phi}\, \hat{\bm{s}}_{k|k}, \\
\label{eq:kf_predict}
\bm{\Sigma}_{k+1|k} &= \bm{\Phi}\, \bm{\Sigma}_{k|k}\, \bm{\Phi}^\top + \bm{Q}_d .
}
Finally, the estimated posterior field on the grid is
\seqn{
\label{eq:fkk}
    \hat{\bm{f}}_{k|k} & =\sqrt{\bm{K}_{gg}} \bm{C}\hat{\bm{s}}_{k|k},\\
    \label{eq:pikk}
    \bm{\Pi}_{k|k} &= \sqrt{\bm{K}_{gg}}\bm{C} \bm{\Sigma}_{k|k} \bm{C}^\top \sqrt{\bm{K}_{gg}}.
}
For any symmetric matrix $X \in \mathbb{R}^{n \times n}$, define the linear operator
\eqn{
\label{eq:pi_operator}
    \mathcal{T}(X) := \bm{K}_{gg}^{1/2} \bm{C} X \bm{C}^{\top} \bm{K}_{gg}^{1/2}. 
}
For brevity, we adopt the shorthand $\bm{\Sigma}_k := \bm{\Sigma}_{k|k-1}$ and $\bm{\Pi}_k := \mathcal{T}(\bm{\Sigma}_k)=\bm{\Pi}_{k|k-1}$. We also write $\hat{\bm{f}}_k := \hat{\bm{f}}_{k|k}$ for the posterior field estimate.
Applying the \ac{KF} recursion in Eq.~\eqref{eq:kf_update} and Eq.~\eqref{eq:kf_predict} to this model yields the \ac{STGPKF}.

\begin{figure*}[t]
    \centering
    \begin{minipage}[c]{0.7\linewidth}
        \centering
        \includegraphics[width=\linewidth]{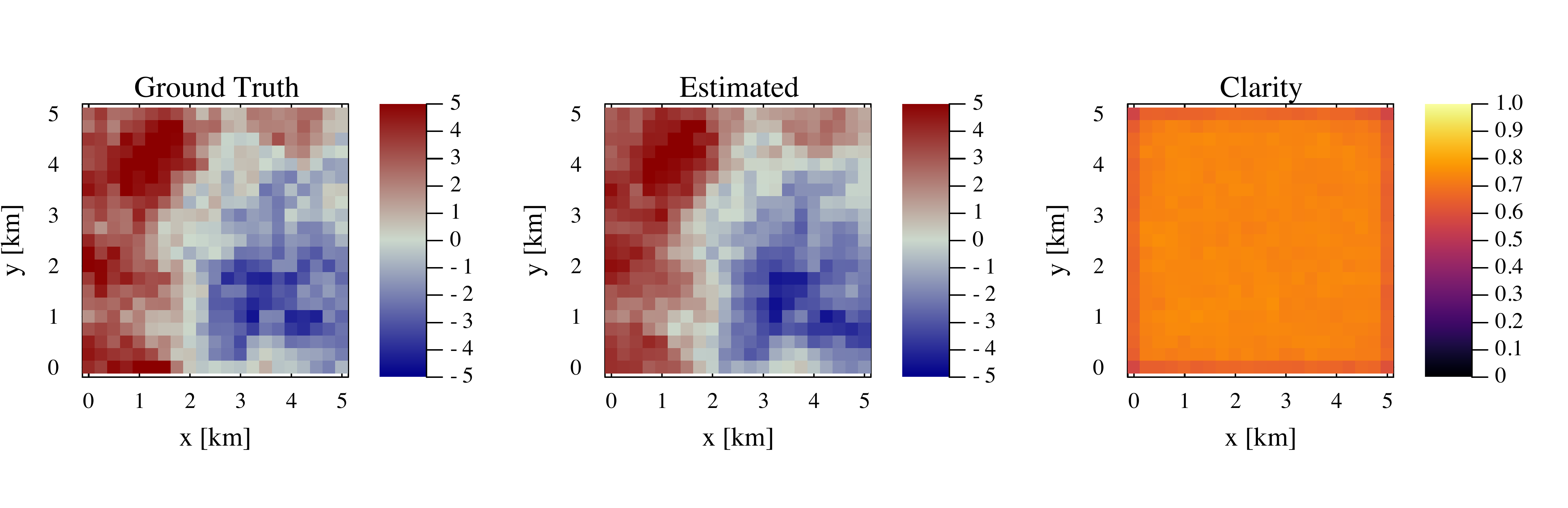}\\[-1mm]
        \small (a)
    \end{minipage}\hfill
    \begin{minipage}[c]{0.30\linewidth}
        \centering
        \includegraphics[width=\linewidth]{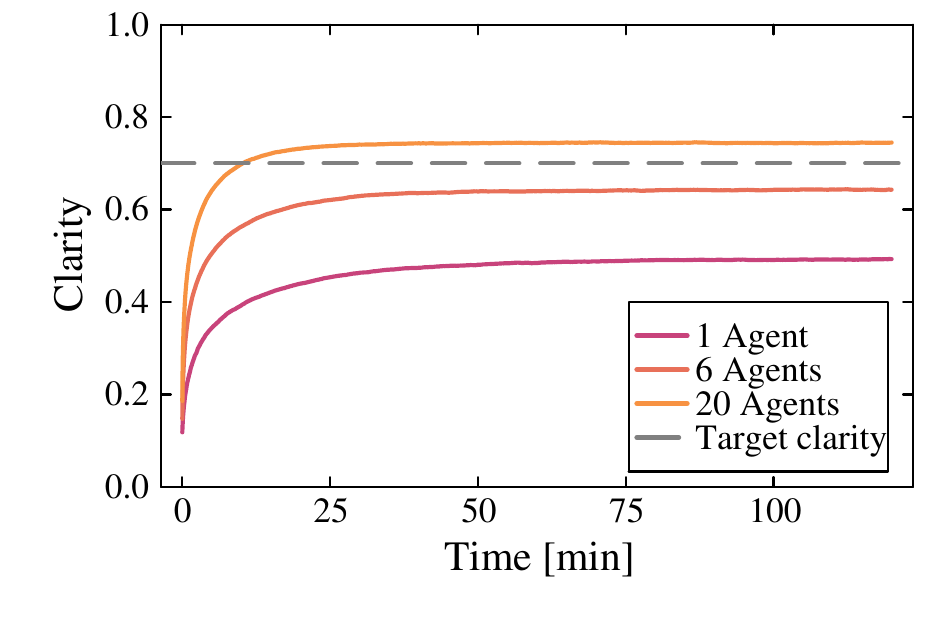}\\[-1mm]
        \small (b)
    \end{minipage}
    \caption{\ac{STGPKF} field-estimation results. (a) Wind field reconstruction and clarity assessment for a single realization with spatial grid spacing $\delta = 0.25\,\mathrm{km}$ and $N_r = 20$. The red–blue color map denotes wind flow direction, and the color intensity represents the wind speed magnitude. (b) Spatially averaged clarity $\bar{q}_k$ over time for different numbers of agents, with one representative randomized sensor-deployment realization shown for each agent count.}
    \label{fig:estimation_results}
\end{figure*}

\subsection{Estimation Performance}
\label{sec:clarity}
To evaluate the estimation performance, we adopt clarity \cite{agrawal2024multi},
which rescales the differential entropy of a continuous random variable onto the range $[0,1]$, with higher values indicating greater certainty.
The differential entropy $h[\bm{Y}]$ of a continuous random variable $\bm{Y} \in \mathbb{R}^n$ is defined as:
$
% h[\bm{Y}] = - \int_{\bm{S}} \rho(x)\,\log \rho(x)\, dx,
h[\bm{Y}] = - \int \rho(x)\,\log \rho(x)\, dx.
$
% where $\bm{S}$ is the support of $\bm{Y}$. 
The clarity of $\bm{Y}$, denoted $q[\bm{Y}]$, is then defined as
\(q[\bm{Y}] = \left( 1 + \frac{e^{2 h[\bm{Y}]}}{(2\pi e)^n} \right)^{-1}\).
Higher clarity values correspond to lower uncertainty, while lower values indicate higher uncertainty.
For a scalar Gaussian random variable $Y \sim \mathcal{N}(\mu,\sigma^2)$, the clarity simplifies to $q[Y] = \frac{1}{1 + \sigma^2}.$
For a Gaussian vector $\bm{Y} \sim \mathcal{N}(0,\bm{\Pi})$ with covariance matrix $\Pi$, we define the marginal clarity for each component using its variance as
\(q_i = \frac{1}{1 + \bm{\Pi}_{ii}}\),
where $\bm{\Pi}_{ii}$ denotes the variance at location $i$.

\subsection{Numerical Evaluation of \ac{STGPKF} and Clarity}

\begin{table}[ht]
\centering
\caption{Simulation parameters.}
\label{tab:sim_params}
\begin{tabular}{l c c}
\hline
\textbf{Parameter} & \textbf{Symbol / Value} & \textbf{Unit} \\
\hline
Temporal kernel std. & $\sigma_t = 2.0$ & km/min \\
Temporal length scale    & $l_t = 60$ & min \\
Spatial kernel std.  & $\sigma_s = 1.0$ & km/min \\
Spatial length scale     & $l_s = 2.0$ & km \\
Spatial resolution       & $\delta = 0.5$ & km \\
Spatial domain           & $[0,5] \times [0,5]$ & km$^2$ \\
Temporal resolution      & $\Delta t = 0.02$ & min \\
Simulation horizon       & $t_{\max} = 2\times60$ & min \\
\begin{editone}Noise intensity\end{editone}
& \begin{editone}$\sigma_c^2 = 0.2$\end{editone}
& \begin{editone}km$^2$/min\end{editone} \\
Number of agents         & $N_r = 1$ & -- \\
\hline
\end{tabular}
\end{table}

% In this section, 
We validate the \ac{STGPKF} and assess its estimation performance using numerical simulations. 
The spatiotemporal field, modeled as a \ac{GP} with temporal and spatial Matérn-$\tfrac{1}{2}$ correlations, admits the following state-space representation~\cite{agrawal2024multi}:
$
\bm{A}_0 = \bmat{-\frac{1}{l_t}},\,
\bm{B}_0 = \bmat{1}, \,
\bm{C}_0 = \bmat{\sigma_t \sqrt{ \frac{2}{l_t} }}
$
Unless otherwise stated, the parameters used in all subsequent simulations are listed in Table~\ref{tab:sim_params}.

Figure~\ref{fig:estimation_results}(a) illustrates a representative realization of sensor deployment, where the agent is randomly placed at each time step.
The ground-truth field represents the underlying wind flow, where the red–blue color map encodes the wind velocity and the color intensity corresponds to the wind speed magnitude.
The \ac{STGPKF} framework effectively reconstructs the dynamic wind field, accurately capturing both the direction and magnitude of the true flow. The clarity map further verifies the estimation reliability: higher clarity values indicate lower uncertainty and greater confidence in the reconstructed field.

To evaluate the information content of the entire field, rather than at individual points, we consider the spatially averaged clarity.
For a given sensor realization, the spatially averaged clarity at time~$t_k$ is defined as
\eqn{
\label{eq:clarity_each_realization}
\bar{q}_k := \frac{1}{N_g}\sum_{i=1}^{N_g} \frac{1}{1 + \bm{\Pi}_{ii,k}}.
}
Here, $\bm{\Pi}_k= \mathcal{T}(\bm{\Sigma}_k)$ denotes the covariance of field-state estimation error.

Figure~\ref{fig:estimation_results}(b) reports how the spatially averaged clarity $\bar q_k$ defined in Eq.~\eqref{eq:clarity_each_realization} evolves over time.
As the number of agents increases, averaged clarity improves monotonically. For example, simulations indicate that deploying 20 agents reaches the target clarity level of 0.7 in steady state. In the following section, instead of relying on simulations to determine the attainable clarity for a given number of sensors, we replace such empirical evaluation with a closed-form upper bound, which allows direct computation of the minimum number of sensors required to meet a prescribed clarity threshold.

\subsection{Steady-State Performance Analysis for \ac{STGPKF}}
A central design question in spatiotemporal field estimation is the following:
\begin{editone}\emph{Given the environmental parameters, how should the number and sensing capabilities of the sensors be selected for a desired estimation-accuracy target prior to deployment?}\end{editone}
To address this question, our first objective is to analyze the expected estimation covariance under randomized sensing.
Specifically, this section develops a steady-state lower bound on the \begin{editone}expected spatially averaged clarity\end{editone} of the continuous-time model, providing a tractable basis for analyzing fundamental performance limits.

\subsubsection{Discrete-Time Upper Bound for \ac{STGPKF}}
We first extend the discrete-time stochastic sensor selection framework of~\cite{gupta2006stochastic}, which derives upper and lower bounds on the expected estimation covariance using the Riccati recursion and demonstrates their tightness in steady state. Although their formulation considers low-dimensional systems with single-sensor activation, its principles remain applicable. Here, we generalize this framework to the \ac{STGPKF} setting, enabling analysis under multi-agent randomized sensing.

For notational convenience, define the discrete-time Riccati operator
parameterized by $(\bm{H},\bm{V})$:
\eqnOne{
\label{eq:riccati_op_disc}
\mathcal{R}_{\bm{H},\bm{V}}(\bm{\Sigma})
&:= \bm{\Phi} \bm{\Sigma} \bm{\Phi}^\top + \bm{Q}_d \\
&- \bm{\Phi} \bm{\Sigma} \bm{H}^\top (\bm{H} \bm{\Sigma} \bm{H}^\top + \bm{V})^{-1} \bm{H} \bm{\Sigma} \bm{\Phi}^\top,
}
where $ \bm{\Sigma} \in \pd^n$.
The Kalman covariance recursion in Eq.~\eqref{eq:kf_update}–\eqref{eq:kf_predict} can be written compactly as
\eqn{
\bm{\Sigma}_{k+1} \;=\; \mathcal{R}_{\bm{H}_k,\bm{V}_k}(\bm{\Sigma}_k).
}
Under randomized sensing, $\bm\Sigma_{k+1}$ is a random matrix. We therefore consider its expectation:
\eqn{
    \bar{\bm{\Sigma}}_{k+1} := \E[\bm{\Sigma}_{k+1}] = \E[ \mathcal{R}_{\bm{H}_k,\bm{V}_k}(\bm{\Sigma}_{k})].
}
For each time step $k$, this expectation can be approximated via Monte Carlo sampling using $N$ independent sensor deployments
$\{(\bm{H}^{(i)}_k,\bm{V}^{(i)}_k)\}_{i=1}^N$.
If deployment $i$ occurs at step $k$ with probability $\pi_i^k$, then
\eqn{
    \E[\bm{\Sigma}_{k+1}]  \approx  \sum_i^{N} \pi^k_i \bm{\Sigma}^{(i)}_{k+1}.
}
However, explicitly evaluating this expectation is intractable, so we seek an upper bound. The following recursion, adapted from Gupta et al.~\cite{gupta2006stochastic}, provides such a result.
\begin{prop} \textbf{{\cite[Thm.~3]{gupta2006stochastic}} [Discrete-time Upper Bound]} 
\label{prop:discrete_upper_bound}
Let $N$ be the number of sensor deployments. If deployments are selected independently across time, with the $i$th deployment chosen at time step $k$ with probability $\pi^k_i$, then the expected covariance is upper bounded by $\bm{\Delta}_{k+1}$, where $\bm{\Delta}_{k}$ is given by the recursion:
\eqnOne{
&\bm{\Delta}_{k+1}  = \sum_{i=1}^{N} \pi_i^k \mathcal{R}_{\bm{H}^{(i)}_k,\bm{V}^{(i)}_k}(\bm{\Delta}_{k})  = \bm{Q}_d + \bm{\Phi} \bm{\Delta}_{k} \bm{\Phi}^\top - \\
& \sum_{i=1}^N \pi^k_i \bm{\Phi}\bm{\Delta}_{k} \!(\bm{H}^{(i)}_k)^\top \!(\bm{V}^{(i)}_k \!+ \!\bm{H}^{(i)}_k \bm{\Delta}_{k}(\bm{H}^{(i)}_k)^\top \!)^{-1} \!\bm{H}^{(i)}_k \!\bm{\Delta}_{k} \bm{\Phi}^\top
}
with the initial condition $\bm{\Delta}_0=\bm{\Sigma}_0$. 
\end{prop}

\begin{pf}
See~\cite[Thm.~3]{gupta2006stochastic}.
\end{pf}

\begin{rem}
    Although rigorous, this bound is recursive and computationally expensive, providing certification but limited design insight. In particular, the absence of a closed-form steady-state solution prevents us from addressing questions such as how performance improves with additional sensors. This limitation motivates the continuous-time analysis in the next section, where explicit closed-form performance bounds can be derived.
\end{rem}

\subsubsection{Continuous-time Upper Bound for \ac{STGPKF}}

To overcome the limitations of the discrete-time recursion, 
we now turn to a continuous-time formulation. 
\begin{editone}To construct a continuous-time approximation of the discrete-time \ac{STGPKF}, the sensor configuration sampled at \(t_k\) is held constant over the corresponding sampling interval, i.e., \(\bm r(t)=\bm r_k\) for \(t\in[t_k,t_{k+1})\).\end{editone}
Following the approach in~\cite{lewis2017optimal}, we express the continuous-time measurement model corresponding to the discrete formulation in Eq.~\eqref{eq:measurement_model} as
\eqn{
\label{eq:measurement_model_continuous}
\bm{y}(t) = \bm{H}(\bm{r}(t)) \bm{s}(t) + \bm{v}(t), \, \bm{v}(t)\sim \mathcal{N} \big(\bm{0},\bm{V}_c(\bm{r}(t))\big),
}
with $\bm{H}(t)=\bm{H}_k$ and $\bm{V}_c(t)=\Delta t\,\bm{V}_k$.
% \edit{Using $\sigma_m^2=\sigma_c^2/\Delta t$ in Eq.~\eqref{eq:measurement_model}, for $t\in[t_k,t_{k+1})$ this covariance can be written as}
% \begingroup
% \color{blue}
% \eqn{
% \bm{V}_c(t)=\sigma_c^2\bm{I}_{N_r}+\Delta t\!\left[
% \bm{K}_{rr}(\bm{r}_k)-\bm{K}_{rg}(\bm{r}_k)\bm{K}_{gg}^{-1}\bm{K}_{gr}(\bm{r}_k)
% \right].
% }
% \endgroup
% \edit{For grid-point sensing, the bracketed term vanishes by Lemma~\ref{lem:K_cancels}, and hence $\bm{V}_c(t)=\sigma_c^2\bm{I}_{N_r}$ is independent of $\Delta t$.}
Together with the continuous-time process model in Eq.~\eqref{eq:process_model_continuous}, this measurement model yields the Kalman–Bucy Riccati differential equation for the covariance in Eq.~\eqref{eq:continuous_riccati}.
 % ~\cite{bucy2005filtering}. 
\begin{editone}Theorems~\ref{thm:continuous_time_upper_bound} and~\ref{thm:analystical_sol_continuous_riccati} characterize the continuous-time solution $\bm{\Delta}(t)$. The following proposition establishes that the expected covariance of the discrete-time \ac{STGPKF} converges to this solution over finite time horizons as $\Delta t\to0$.\end{editone}

\begin{editone}
Although the general measurement model in Sec.~5.2 permits arbitrary off-grid sensing locations, the finite-horizon convergence analysis below specializes to on-grid sensing.

\begin{lem}
\label{lem:K_cancels}
If sensors take measurements at the grid points (i.e., sensing locations coincide with the discretization grid), then
\eqn{
\bm{K}_{rr} - \bm{K}_{rg} \bm{K}_{gg}^{-1} \bm{K}_{gr} = 0.
}
\end{lem}
\begin{pf}
    See Appendix~\ref{appendix:K_cancels}.
\end{pf}

\begin{prop}
    %  \textbf{[Finite-Horizon Continuous-Time Limit]}
\label{prop:stgpkf_continuous_limit}
Under the i.i.d. sampling model specified above and the 
condition of Lemma~\ref{lem:K_cancels}, consider the limit $\Delta t\to0$ with all other model parameters held fixed. Assume that the initial covariance $\bm{\Sigma}_0\in\pd^n$ is deterministic and independent of the sensing process. 
Let $\bm{\Sigma}_k$ be generated by the discrete-time \ac{KF} in Eqs.~\eqref{eq:kf_update}-\eqref{eq:kf_predict}, and let $\bm{\Delta}(t)$ solve Eq.~\eqref{eq:Delta} with $\bm{\Delta}(0)=\bm{\Sigma}_0$.
% and
% $
% \bar{\bm{G}}=\E\!\left[\bm{H}_k^\top(\sigma_c^2\bm{I}_{N_r})^{-1}\bm{H}_k\right].
% $
Then, for every finite $T>0$,
\eqn{
\label{eq:stgpkf_continuous_limit}
\lim_{\Delta t\to0}\max_{k:\,t_k\leq T}
\left\|\E[\bm{\Sigma}_k]-\bm{\Delta}(t_k)\right\|_F=0.
}
\end{prop}
\begin{pf}
See Appendix~\ref{appendix:stgpkf_continuous_limit}.
\end{pf}

\begin{cor}
    % \textbf{[Discrete-Bound Continuous-Time Limit]}
\label{cor:discrete_bound_continuous_limit}
Under the assumptions of Proposition~\ref{prop:stgpkf_continuous_limit}, let
$\bm{\Delta}_k$ be the discrete-time upper-bound sequence in
Proposition~\ref{prop:discrete_upper_bound}, with $\bm{\Delta}_0=\bm{\Sigma}_0$. Then, for every finite $T>0$,
\eqn{
\label{eq:discrete_bound_continuous_limit}
\lim_{\Delta t\to0}\max_{k:\,t_k\leq T}
\left\|\bm{\Delta}_k-\bm{\Delta}(t_k)\right\|_F=0.
}
\end{cor}
% \begin{pf}
% See Appendix~\ref{appendix:discrete_bound_continuous_limit}.
% \end{pf}
\end{editone}

\begin{rem}
\label{remark:continuous}
\begin{editone}
% Proposition~\ref{prop:stgpkf_continuous_limit} shows that, on every fixed finite time horizon, the expected covariance of the discrete-time \ac{STGPKF} converges uniformly to the continuous-time solution \(\bm{\Delta}(t)\) as \(\Delta t\to0\).
Proposition~\ref{prop:stgpkf_continuous_limit} and Corollary~\ref{cor:discrete_bound_continuous_limit} show that, on every fixed finite time horizon, both the expected covariance \(\E[\bm{\Sigma}_k]\) and the discrete-time upper bound \(\bm{\Delta}_k\) converge uniformly to the continuous-time solution \(\bm{\Delta}(t_k)\) as \(\Delta t\to0\).
\end{editone}
Beyond establishing this connection, the continuous-time formulation admits a closed-form steady-state solution for the covariance upper bound, enables a characterization of the grid-independent limit of the associated clarity lower bound as the spatial grid is refined, and reveals the sensing quantity governing this limit. We show these properties in the next section.
% Moreover, the numerical results below illustrate the corresponding finite-horizon convergence through a 
% \begin{editone}spatially averaged covariance metric\end{editone}.
\end{rem}

% \subsection{Simulations: Continuous vs. Discrete Upper Bound}

% \begin{figure}
%     \centering
%     \includegraphics[width=0.9\linewidth]{figure/continuous_expected.pdf}
%     \caption{
%     Comparison of the evolution of the expected covariance $\E[\bm{\Sigma}_k]$ and its upper bounds, $\bm{\Delta}_k$ (discrete time) and $\bm{\Delta}(t)$ (continuous time), with a sampling interval of $\Delta t = 0.05$.    The metric $\mathcal{L}(\cdot)$ denotes the linear operator that maps a covariance matrix to the mean of its diagonal entries. \todo{to be discussed}
%      }
%     \label{fig:bound_compare}
% \end{figure}

\begin{figure}[t]
    \centering
    \includegraphics[width=\linewidth]{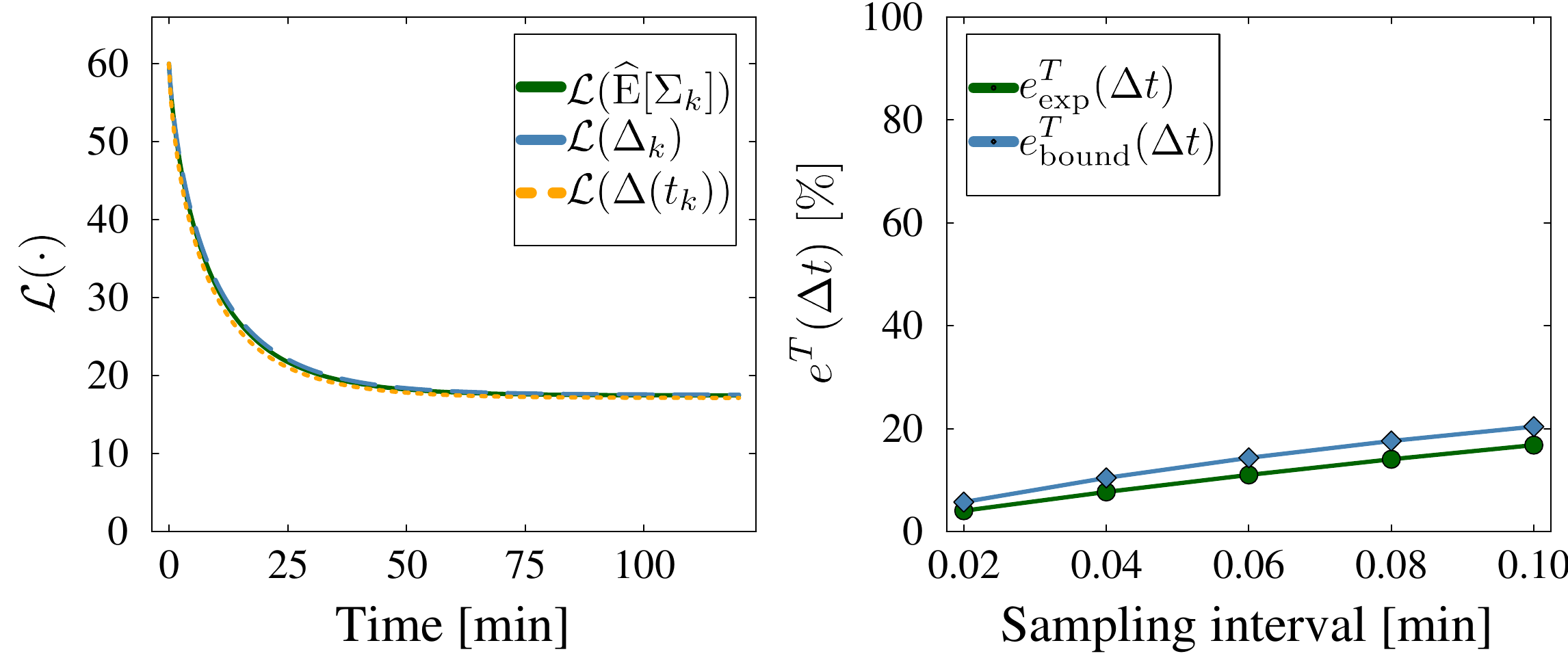}
    \caption{\begin{editone}Covariance comparison ($T=120\,\mathrm{min}$): the spatially averaged covariance metric $\mathcal L(\cdot)$ evaluated for the Monte Carlo covariance estimate and the discrete- and continuous-time bounds at $\Delta t=0.02\,\mathrm{min}$ (left), and maximum relative errors in Eqs.~\eqref{eq:empirical_relative_linear_error}--\eqref{eq:bound_relative_linear_error} versus $\Delta t$ (right).\end{editone}}
    \label{fig:continuous_vs_discrete}
\end{figure}

\begin{editone}To evaluate this approximation, we run the discrete-time \ac{STGPKF} for $M=30$ Monte Carlo trials with the sensor location independently resampled at every step.
Let $\widehat{\E}[\bm{\Sigma}_k]:=\frac{1}{M}\sum_{m=1}^{M}\bm{\Sigma}_k^{(m)}$
denote the Monte Carlo estimate of the expected covariance. We compare it with the discrete-time upper bound $\bm{\Delta}_k$ adapted from \cite{gupta2006stochastic} and the continuous-time upper bound $\bm{\Delta}(t_k)$. For visualization, define the spatially averaged covariance metric
$\mathcal L(\bm X):=\operatorname{tr}(\bm X)/N_g$ and the following relative errors:
\eqn{
&e_{\mathrm{exp}}^{T}(\Delta t)
:=\max_{k:\,t_k\leq T}
\frac{\left|
\mathcal L\!\left(\widehat{\E}[\bm{\Sigma}_k]\right)
-\mathcal L\!\left(\bm{\Delta}(t_k)\right)
\right|}
{\left|\mathcal L\!\left(\widehat{\E}[\bm{\Sigma}_k]\right)\right|},
\label{eq:empirical_relative_linear_error}
\\
&e_{\mathrm{bound}}^{T}(\Delta t)
:=\max_{k:\,t_k\leq T}
\frac{\left|
\mathcal L\!\left(\bm{\Delta}_k\right)
-\mathcal L\!\left(\bm{\Delta}(t_k)\right)
\right|}
{\left|\mathcal L\!\left(\bm{\Delta}_k\right)\right|}.
\label{eq:bound_relative_linear_error}
}
\end{editone}
\begin{editone}Figure~\ref{fig:continuous_vs_discrete} illustrates the convergence behavior implied by Proposition~\ref{prop:stgpkf_continuous_limit} and Corollary~\ref{cor:discrete_bound_continuous_limit}. Both relative errors in the spatially averaged covariance metric decrease as \(\Delta t\) decreases, showing that the continuous-time approximation becomes more accurate as the temporal discretization is refined.
% For finite $\Delta t$, the fidelity of the continuous-time surrogate depends on using a sufficiently fine temporal discretization.
\end{editone}

\begin{editone}
\subsubsection{Continuous-time Lower Bound of Expected Spatially Averaged Clarity}

For each sensing realization, let the spatially averaged clarity be
\eqn{
\label{eq:continuous_spatial_mean_clarity}
\bar q(t):=\frac{1}{N_g}\sum_{i=1}^{N_g}\frac{1}{1+\bm{\Pi}_{ii}(t)}.
}
Under stochastic sensing, the expected spatially averaged clarity is therefore
\eqn{
\label{eq:expected_clarity}
\E[\bar q(t)]
:=\frac{1}{N_g}\sum_{i=1}^{N_g}
\E\!\left[\frac{1}{1+\bm{\Pi}_{ii}(t)}\right].
}
% Here $\E[\bm{\Pi}(t)] =\mathcal{T}(\E[\bm{\Sigma}(t)])$ denotes the expected covariance of the field state. 
% We refer to
% \[
% \bar q_{\E[\Pi]}(t):=\frac{1}{N_g}\sum_{i=1}^{N_g}
% \frac{1}{1+\E[\bm\Pi_{ii}(t)]}
% \]
% as the spatially averaged clarity evaluated at the expected covariance. 

% Building on the covariance bound established earlier, we now seek a corresponding relation for the expected clarity in Eq.~\eqref{eq:expected_clarity}.

\begin{thm} \textbf{[Lower Bound on Expected Spatially Averaged Clarity]}
\label{thm:avg_clarity}
% Consider the Matérn-$\tfrac{1}{2}$ kernel ($n_k=1$), $\bm{C}_0$ reduce to a scalar, i.e., $C = \bm{C}_0 I$ with $\bm{C}_0 >0$.
Suppose $\bm{\Delta}(t)\succeq \E[\bm{\Sigma}(t)]$ for all $t\ge t_0$.
Then, the expected spatially averaged clarity admits the lower bound
\eqn{
\label{eq:q_delta}
% \bar q_{\bm{\Delta}^{\Pi}(t)} := \frac{1}{N_g}\sum_{i=1}^{N_g} \frac{1}{1 + \bm{\Delta}^{\Pi}_{ii}(t)},}
\bar q_{\Delta^{\Pi}(t)} :=  \frac{1}{1 + \frac{1}{N_g}\operatorname{tr}(\bm{\Delta}^{\Pi}(t))},
}
such that, for every $t\ge t_0$,
\eqn{
\label{eq:expected_clarity_large_q_delta}
\E[\bar q(t)] \;\ge\; \bar q_{\Delta^{\Pi}(t)}.
}
Here, $\bm{\Delta}^{\Pi}(t) = \mathcal{T}\big(\bm{\Delta}(t)\big)$
is an upper bound on the expected covariance of the field state $\bm{f}$.
\end{thm}

\begin{pf}
\begin{editone}
Since $\bm{\Delta}(t)-\E[\bm{\Sigma}(t)]\succeq\bm 0$ and Eq.~\eqref{eq:pi_operator}
maps positive semidefinite matrices to positive semidefinite matrices,
\eqnN{
\mathcal{T}(\bm{\Delta}(t))-\mathcal{T}(\E[\bm{\Sigma}(t)])
=\mathcal{T}\big(\bm{\Delta}(t)-\E[\bm{\Sigma}(t)]\big)\succeq\bm 0.
}
Hence, $\E[\bm{\Pi}(t)]=\mathcal{T}(\E[\bm{\Sigma}(t)])
\preceq\mathcal{T}(\bm{\Delta}(t))=\bm{\Delta}^{\Pi}(t)$.
\end{editone}

Let $f_q(x)=\frac{1}{1+x}$. Since $f_q$ is convex and strictly decreasing for $x\ge0$, Jensen's inequality and the covariance bound give
\seqn{
 \E[\bar q(t)]
%  &=\frac{1}{N_g}\sum_{i=1}^{N_g}\E\!\left[f_q(\bm{\Pi}_{ii}(t))\right] \\
 &\ge \frac{1}{N_g}\sum_{i=1}^{N_g} f_q \big(\E[\bm{\Pi}_{ii}(t)]\big) \label{eq:clarity_jensen_expectation} \\
 & \ge \frac{1}{N_g}\sum_{i=1}^{N_g} f_q(\bm{\Delta}^{\Pi}_{ii}(t)) \label{eq:1_clarity} \\
 &\ge  f_q(\frac{1}{N_g}\sum_{i=1}^{N_g} \bm{\Delta}^{\Pi}_{ii}(t))  \label{eq:2_clarity} \\
 & = f_q(\frac{1}{N_g}\operatorname{tr} (\bm{\Delta}^{\Pi}(t)))\label{eq:3_clarity}  
}
Eq.~\eqref{eq:clarity_jensen_expectation} follows from Jensen's inequality applied to each random diagonal entry $\bm{\Pi}_{ii}(t)$. Eq.~\eqref{eq:1_clarity} follows from the monotonicity of $f_q$ and the relation
$\bm{\Delta}^{\Pi}(t) \succeq \E[\bm{\Pi}(t)]$, which implies
$\bm{\Delta}^{\Pi}_{ii}(t) \ge \E[\bm{\Pi}_{ii}(t)]$ for all $i\in\{1,\dots,N_g\}$.  
Eq.~\eqref{eq:2_clarity} follows from Jensen's inequality across the spatial locations. Eq.~\eqref{eq:3_clarity} uses the definition of the matrix trace,
$\operatorname{tr}(\bm{\Delta}^{\Pi}(t)) = \sum_{i=1}^{N_g}\bm{\Delta}^{\Pi}_{ii}(t)$.
\end{pf}

\begin{editone}Under the conditions of Theorem~\ref{thm:analystical_sol_continuous_riccati}, $\bm{\Delta}(t)\to\bm{\Delta}_\infty$. Since $\mathcal T$ is linear,
$\bm{\Delta}^{\Pi}(t)\to\bm{\Delta}^{\Pi}_\infty:=\mathcal T(\bm{\Delta}_\infty)$.
Define\end{editone}
\eqn{
\label{eq:clarity_steady_state}
\bar{q}_{\Delta^{\Pi}_\infty} = \frac{1}{1 + \frac{1}{N_g}\operatorname{tr}(\bm{\Delta}^{\Pi}_{\infty})}.
}
\begin{editone}
Together with Eq.~\eqref{eq:expected_clarity_large_q_delta}, this gives
\eqnN{
\liminf_{t\to\infty}\E[\bar q(t)]
&\geq \liminf_{t\to\infty}\bar q_{\Delta^{\Pi}(t)}=\bar q_{\Delta^{\Pi}_\infty}.
}
Thus, $\bar q_{\Delta^{\Pi}_\infty}$ is an asymptotic lower bound on the expected spatially averaged clarity. \end{editone}
As $\bm{\Delta}^{\Pi}_\infty$ depends on the eigenvalues of the averaged information matrix~$\bar{\bm{G}}$ by Theorem~\ref{thm:analystical_sol_continuous_riccati}, we next examine the structure of~$\bar{\bm{G}}$.
\end{editone}

\subsection{Analytical Characterization and Fundamental Limits}
\label{sec:asymptotic}
Building on the steady-state clarity lower bound derived in the previous section, this section addresses the second objective: to characterize the fundamental limits of this bound under stochastic sensing.

First, we derive a closed-form expression for the averaged information matrix $\bar{\bm{G}}$.
Then, motivated by prior work~\cite{tzoumas2016sensor}, which investigates how system size affects estimation performance, we analyze the structure of the averaged information matrix $\bar{\bm{G}}$ to identify the factors governing this bound.
We aim to show that varying the number of grid points within the same compact domain does not fundamentally affect the steady-state clarity lower bound; rather, the bound is determined by intrinsic sensing parameters.
As the grid is refined ($\delta \to 0$), the steady-state lower bound on \begin{editone}expected spatially averaged clarity\end{editone} $\bar{q}_{\Delta^{\Pi}_\infty}$ becomes independent of discretization and, for a given temporal kernel, depends mainly on the number of agents, measurement noise, and sampling interval.

\subsubsection{Closed-form of Averaged Information Matrix}

\begin{lem}
\label{lem:kgrkrg}
Let $P_g$ be as defined in Eq.~\eqref{eq:x_grid}
and suppose $N_r$ agents \begin{editone}independently and uniformly\end{editone} choose locations from $P_g$. Then,
 $\E[ \bm{K}_{gr} \bm{K}_{rg}] = \frac{N_r}{N_g} \bm{K}_{gg}^2.$
\end{lem}
\begin{pf}
    See Appendix~\ref{appendix:kgrkrg}.
\end{pf}

Combining Lemma~\ref{lem:K_cancels} and Lemma~\ref{lem:kgrkrg}, we obtain the following expression for the averaged gain matrix $\bar{\bm{G}}$:

\begin{lem}
\label{lem:G_bar}
Define the sensing parameter 
\begin{editone}\(\theta := \frac{N_r}{\sigma_m^2\,\Delta t}=\frac{N_r}{\sigma_c^2}\).\end{editone}
% \edit{which is the number of sensing agents normalized by the continuous-time measurement-noise intensity.}
Under the same assumptions as in Lemma~\ref{lem:K_cancels} and Lemma~\ref{lem:kgrkrg}, for $N_g$ grid points and $N_r$ agents, the averaged information matrix has the following expression:
\begin{editone}
\eqnOne{
\label{eq:g_bar_closed}
\bar{\bm{G}} = \E[\bm{H}^\top \bm{V}_c^{-1} \bm{H}]
%  = \frac{N_r}{\sigma_c^2}  \bm{C}^\top \frac{\bm{K}_{gg}}{N_g} \bm{C} 
 = \theta \bm{C}^\top \frac{\bm{K}_{gg}}{N_g} \bm{C} 
}
\end{editone}
\end{lem}
\begin{pf}See Appendix~\ref{appendix:G_bar}.
\end{pf}

% \begin{rem}
It is observed that the averaged information matrix $\bar{\bm{G}}$ naturally separates the influence of the sensing configuration from that of the system dynamics. 
\begin{editone}Therefore, increasing $N_r$ or decreasing $\sigma_c^2$ increases $\bar{\bm G}$. Since $\sigma_c^2=\sigma_m^2\Delta t$, decreasing either $\sigma_m^2$ or $\Delta t$ while holding the other fixed has the same effect.\end{editone}
In contrast, the matrix term $\bm{C}^\top \frac{\bm{K}_{gg}}{N_g} \bm{C} $ encapsulates the spatial correlation structure of the environment.
Moreover, since~$\bar{\bm{G}}$ is mainly determined by the spatial kernel matrix~$\bm{K}_{gg}$, its spectral properties govern how sensing resources affect estimation clarity.
This observation motivates a spectral analysis of~$\bm{K}_{gg}$ in the next subsection.

% \end{rem}

\subsubsection{Fundamental Performance Limits}
\label{sec:design_problem}
In this section, we investigate how the system size, represented by the number of grid points to be estimated, affects the averaged estimation performance.
Specifically, we will describe how the discrete spatial kernel matrix $\bm{K}_{gg}$ behaves as the grid resolution becomes arbitrarily fine.

We first require spectral properties of the kernel integral operator, stated in the following lemma.
\begin{lem}\textbf{[Mercer's theorem]}
\label{lem:mercer}
Consider a continuous, symmetric, positive semi-definite kernel $k_S$ defined in Eq.~\eqref{eq:ks} on a compact domain $D$.
Assume the operator $T$ takes a function $f(\cdot)$ as its argument and outputs a new function as:
\eqn{
\label{eq:T_operator}
T f(\bm{p}) := \int_D k_S(\bm{p},\bm{p}') f(\bm{p}') \, d\bm{p}',
}
The operator $T$ is called the Hilbert–Schmidt integral operator.
Then, there exists an orthonormal basis $\{\psi_i(\cdot)\}_{i=1}^\infty$ of $L_2(D)$ consisting of eigenfunctions of $T$ such that the corresponding eigenvalue sequence $\{\nu_i\}_{i=1}^\infty$ is nonnegative and $k_S(\bm{p},\bm{p}') = \sum_{i=1}^\infty \nu_i \psi_i(\bm{p}) \psi_i(\bm{p}'),$
where the convergence is absolute and uniform.
Moreover, $T$ is trace class with
\eqn{
\sum_{i=1}^\infty \nu_i = \int_D k_S(\bm{p},\bm{p})\, d\bm{p} <\infty.
}
\end{lem}
\begin{pf}
See~\cite[Thm.~2]{ghojogh2021reproducing} and~\cite[Cor.~5.4,Thm~5.2]{arendt2019operators}.
\end{pf}

Notice that the spatial kernel matrix $\bm{K}_{gg}$ defined in Eq.~\eqref{eq:kgg_delta} is a Nyström approximation of the continuous kernel operator $T$. This approximation is constructed by sampling the continuous kernel function $k_S$ on a uniform spatial grid with spacing $\delta$. The following lemma shows that as $\delta \to 0$, the eigenvalues of discrete operator $T_\delta$ converge to those of $T$.

\begin{lem}
\label{lem:nystrom}
Let $D \subset \R^d$ be a compact domain with finite  Lebesgue measure $|D| > 0$. For a uniform grid spacing $\delta >0$, define the Nyström approximation operator:
\eqn{
(T_{\delta} f)(\bm{p}) := \frac{|D|}{N_g} \sum_{j=1}^{N_g} k_S(\bm{p}, \bm{p}_j) f(\bm{p}_j), N_g = \frac{|D|}{\delta^d}
}
where $\{\bm{p}_j\}_j^{N_g} \subset D$ are grid points and $k_S$ is a continuous, symmetric, positive-definite kernel defined in Eq.~\eqref{eq:ks}.
Let $\{\nu_{i,\delta} \}_{i=1}^{N_g}$ and $\{\nu_{i} \}_{i=1}^{\infty}$ denote eigenvalues of $T_{\delta}$ and of continuous operator $T$ defined in Lemma~\ref{lem:mercer}, respectively.
Then, as $\delta \to 0$ ($N_g \to \infty $), $\nu_{i, \delta} \to \nu_i, \, \forall i \geq 1.
$

\end{lem}
\begin{pf}
 See~\cite[Thm.~4.2]{cai2020eigenvalue}.
\end{pf}

We can now quantify the asymptotic behavior of $\bar{q}_{\Delta^{\Pi}_\infty}$ as the grid becomes dense, which constitutes one of our main results.

\begin{thm}
\label{thm:clarity_converge}
\begin{editone}
Under condition of Lemma~\ref{lem:K_cancels}, and Lemma~\ref{lem:kgrkrg}, and the resulting closed form for $\bar{\bm G}$ in Lemma~\ref{lem:G_bar},
\end{editone}
consider the Matérn-$\tfrac{1}{2}$ kernel ($n_k=1$) with $\bm{C} = C_0 \bm{I}$ and eigenvalues $\{\nu_{i} \}_{i=1}^{\infty}$ of the associated continuous kernel operator in Eq.~\eqref{eq:T_operator} on a given spatial domain.
Define the function $\gamma : \mathbb{R}_+ \to \mathbb{R}_+$ associated with the closed-form Riccati solution in Theorem~\ref{thm:analystical_sol_continuous_riccati} as
\eqn{
\gamma(\lambda) := \frac{-q_c}{a - \sqrt{a^2 + q_c \lambda}}.
}
Then, as the grid spacing $\delta \to 0$, when the environment (kernel parameters and domain) is fixed, the steady-state lower bound on \begin{editone}expected spatially averaged clarity\end{editone} defined in Eq.~\eqref{eq:clarity_steady_state} converges to a finite limit, i.e.,
\eqn{
\lim_{\delta\to 0} \bar{q}_{\Delta^{\Pi}_\infty}
=\frac{1}{1 + C_0^{\,2}\sum_{i=1}^{\infty} \frac{\nu_i}{|D|}  \gamma\bigl(\theta C_0^{2} \frac{\nu_i}{|D|} \bigr)}  
}
which depends on the sensing parameter $\begin{editone}\theta = \frac{N_r}{\sigma_m^2\Delta t}=\frac{N_r}{\sigma_c^2}.\end{editone}$

\end{thm}
\begin{pf}
According to Lemma~\ref{lem:nystrom}, the matrix form of the discrete kernel operator $T_\delta$ is $\frac{|D|\bm{K}_{gg}}{N_g}$. 
 Since $\frac{|D|\bm{K}_{gg}}{N_g} \in \pd^{N_g}$, it admits an eigendecomposition
$
\frac{|D|\bm{K}_{gg}}{N_g} = \bm{U} \bm{\Lambda} \bm{U}^\top,
$
where $\bm{U}$ is an orthogonal matrix. For a discrete spatial domain with grid spacing $\delta$, we have the diagonal matrix $\bm{\Lambda}= \mathrm{diag}(\nu_{1,\delta}, \ldots, \nu_{N_g,\delta})$ as given in Lemma~\ref{lem:nystrom}. Similarly, 
$
\bm{K}_{gg}  = \bm{U} \bm{\Lambda}_K \bm{U}^\top,
$
where $\bm{\Lambda}_K:=\frac{N_g}{|D|}\bm{\Lambda}$.
From Eq.~\eqref{eq:g_bar_closed}, we know $\bar{\bm{G}}$ shares the same orthogonal matrix $\bm{U}$ as $\frac{\bm{K}_{gg}}{N_g}$, i.e.,
\eqnN{
\bar{\bm{G}} 
= \bm{U} \bm{\Lambda}_G \bm{U}^\top,\, \bm{\Lambda}_G=\mathrm{diag}(\lambda_{1,\delta}, \ldots, \lambda_{N_g,\delta})
}
where $\{\lambda_{i,\delta} \}_{i=1}^{N_g}$ are the eigenvalues of $\bar{\bm{G}}$ with each value $\lambda_{i,\delta}  = \theta C_0^2 \frac{\nu_{i,\delta}}{|D|}$. 
Then the eigenvalues $\gamma_{i,\delta}$ of $\bm{\Delta}_{\infty}$ satisfy $\gamma_{i,\delta} = \gamma(\lambda_{i,\delta}) =  \gamma(\theta C_0^2 \frac{\nu_{i,\delta}}{|D|})$.
According to Theorem~\ref{thm:analystical_sol_continuous_riccati}, $\bm{\Delta}_\infty$ also shares the same orthogonal matrix $\bm{U}$ as $\bar{\bm{G}}$, i.e.,
\eqnN{
\bm{\Delta}_\infty &= \bm{U} \bm{\Lambda}_\Delta \bm{U}^{\top},\, \bm{\Lambda}_\Delta= \mathrm{diag}\big(\gamma_{1,\delta}, \ldots, \gamma_{N_g,\delta} \big).
}
The corresponding normalized steady-state field-state covariance bound is
\eqnN{
 \frac{\bm{\Delta}^{\Pi}_\infty}{N_g}
 &= \frac{1}{N_g}\mathcal{T}(\bm{\Delta}_\infty)= \frac{1}{N_g}C_0^2 \bm{K}_{gg}^{1/2} \bm{\Delta}_\infty \bm{K}_{gg}^{1/2} \\
&= \frac{1}{N_g}C_0^2 (\bm{U} \bm{\Lambda}_K^{1/2} \bm{U}^{\top}) 
   (\bm{U} \bm{\Lambda}_{\Delta} \bm{U}^{\top}) 
   (\bm{U} \bm{\Lambda}_K^{1/2} \bm{U}^{\top})\\
% &= \frac{1}{N_g}C_0^2 \bm{U} (\bm{\Lambda}_K^{1/2} \bm{\Lambda}_{\Delta} \bm{\Lambda}_K^{1/2}) \bm{U}^{\top},\\
&= C_0^2 \bm{U} (\frac{\bm{\Lambda}_{K}}{N_g} \bm{\Lambda}_{\Delta} ) \bm{U}^{\top}= C_0^2 \bm{U} (\frac{\bm{\Lambda}}{|D|} \bm{\Lambda}_{\Delta} ) \bm{U}^{\top}.
}
\begin{editone}Therefore, the eigenvalues of $ \frac{\bm{\Delta}^{\Pi}_\infty}{N_g}$ are $\{C_0^2\frac{\nu_{i,\delta}}{|D|} \gamma_{i,\delta} \}_{i=1}^{N_g}$.\end{editone}
Hence,
\eqnN{
\frac{1}{N_g}\operatorname{tr}(\bm{\Delta}^{\Pi}_{\infty}) 
= \sum_{i=1}^{N_g} C_0^2 \frac{\nu_{i,\delta}}{|D|} \gamma_{i,\delta}
= C_0^2 \sum_{i=1}^{N_g}  \frac{\nu_{i,\delta}}{|D|} 
\gamma(\theta C_0^2 \frac{\nu_{i,\delta}}{|D|})
}
By Mercer’s theorem in Lemma~\ref{lem:mercer} and its Nyström approximation in Lemma~\ref{lem:nystrom}, 
$\{\nu_{i,\delta} \}_{i=1}^{N_g}$ are approximations of the eigenvalues $\{\nu_{i} \}_{i=1}^{\infty}$ of the continuous kernel operator, satisfying $\nu_{i,\delta} \to \nu_i$ as $\delta \to 0$
and $\sum_{i=1}^\infty \nu_i <\infty$.
In addition, $\gamma(\cdot)$ is continuous and bounded:
$
0\leq \gamma(\lambda)\leq  \gamma(0)=:C_\gamma <\infty.
$
Hence,
$
0\leq \frac{\nu_i}{|D|} \gamma(\theta C_0^2 \frac{\nu_i}{|D|}) \leq \frac{\nu_i}{|D|} C_\gamma.
$
Therefore,
\eqnN{
\lim_{\delta\to 0} \frac{\operatorname{tr}\bigl(\bm{\Delta}^{\bm{\Pi}}_{\infty}\bigr)}{N_g}\,
&=
C_0^{\,2}\!\sum_{i=1}^{\infty} \frac{\nu_i}{|D|} \gamma\bigl(\theta C_0^{2}\frac{\nu_i}{|D|}  \bigr)\\
&\leq C_0^{\,2} C_\gamma \sum_{i=1}^{\infty} \frac{\nu_i}{|D|}\,.
% < \infty
}
Thus, the limit is finite.
Since $f_q(x) = \frac{1}{1+x}$ is continuous on $[0,\infty)$, continuity gives
\eqnN{
\lim_{\delta\to 0} \bar{q}_{\Delta^{\Pi}_\infty}
&=\frac{1}{1 + \lim_{\delta\to 0} \frac{1}{N_g}\operatorname{tr}(\bm{\Delta}^{\Pi}_{\infty})}  \\
&=\frac{1}{1 + C_0^{\,2}\sum_{i=1}^{\infty} \frac{\nu_i}{|D|}  \gamma\bigl(\theta C_0^{2} \frac{\nu_i}{|D|} \bigr)}\,.
}
When the environment (kernel parameters and domain) is fixed,
the steady-state clarity lower bound converges to a constant that depends solely on the sensing parameter $\theta$, independent of the grid resolution.
\end{pf}

\begin{rem}
Theorem~\ref{thm:clarity_converge} shows that $\bar{q}_{\Delta^{\Pi}_\infty}$ converges as $\delta \to 0$, implying that beyond a certain grid resolution, adding more grid points no longer significantly changes the estimation performance.
\begin{editone}When the temporal kernel parameters $(\bm A_0,\bm B_0,\bm C_0)$ are fixed, the limiting value depends on the number of agents $N_r$, measurement noise variance $\sigma_m^2$, and sampling interval $\Delta t$ only through
$\theta$.\end{editone}
\end{rem}

\begin{figure}[t]
    \centering
    \includegraphics[width=0.9\linewidth]{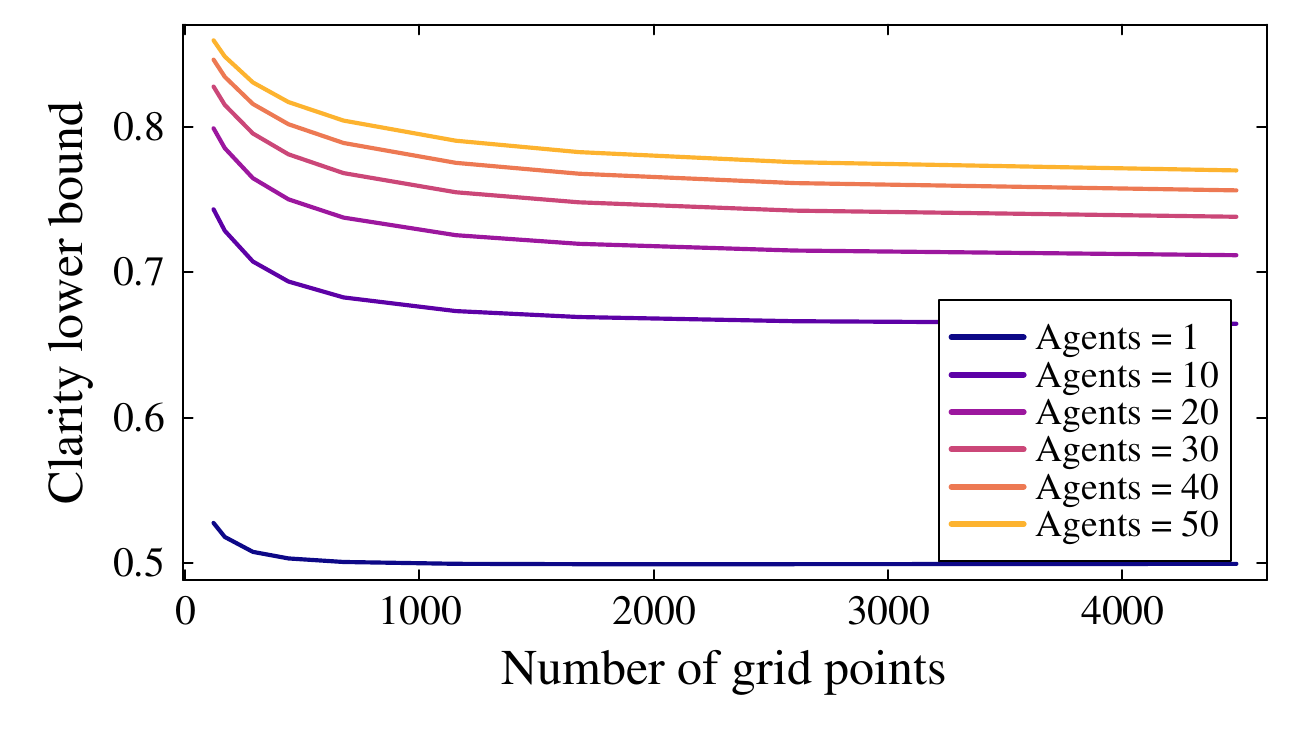}
    \caption{Steady-state lower bound on \begin{editone}expected spatially averaged clarity\end{editone} $\bar{q}_{\Delta^{\Pi}_\infty}$ versus spatial grid size $N_g$. Each curve corresponds to a different $\theta(N_r)$, representing a different number of agents.}
    \label{fig:ng_converge}
\end{figure}

Figure~\ref{fig:ng_converge} empirically validates Theorem~\ref{thm:clarity_converge}. The curves are parameterized by $\theta$, showing that the steady-state clarity lower bound $\bar{q}_{\Delta^{\Pi}_\infty}$ converges to distinct values determined by $\theta$. This confirms that the bound is governed by intrinsic sensing parameters rather than spatial discretization.

\subsubsection{Number of Sensors}

\begin{figure}[t]
    \centering
    \includegraphics[width=0.46\linewidth]{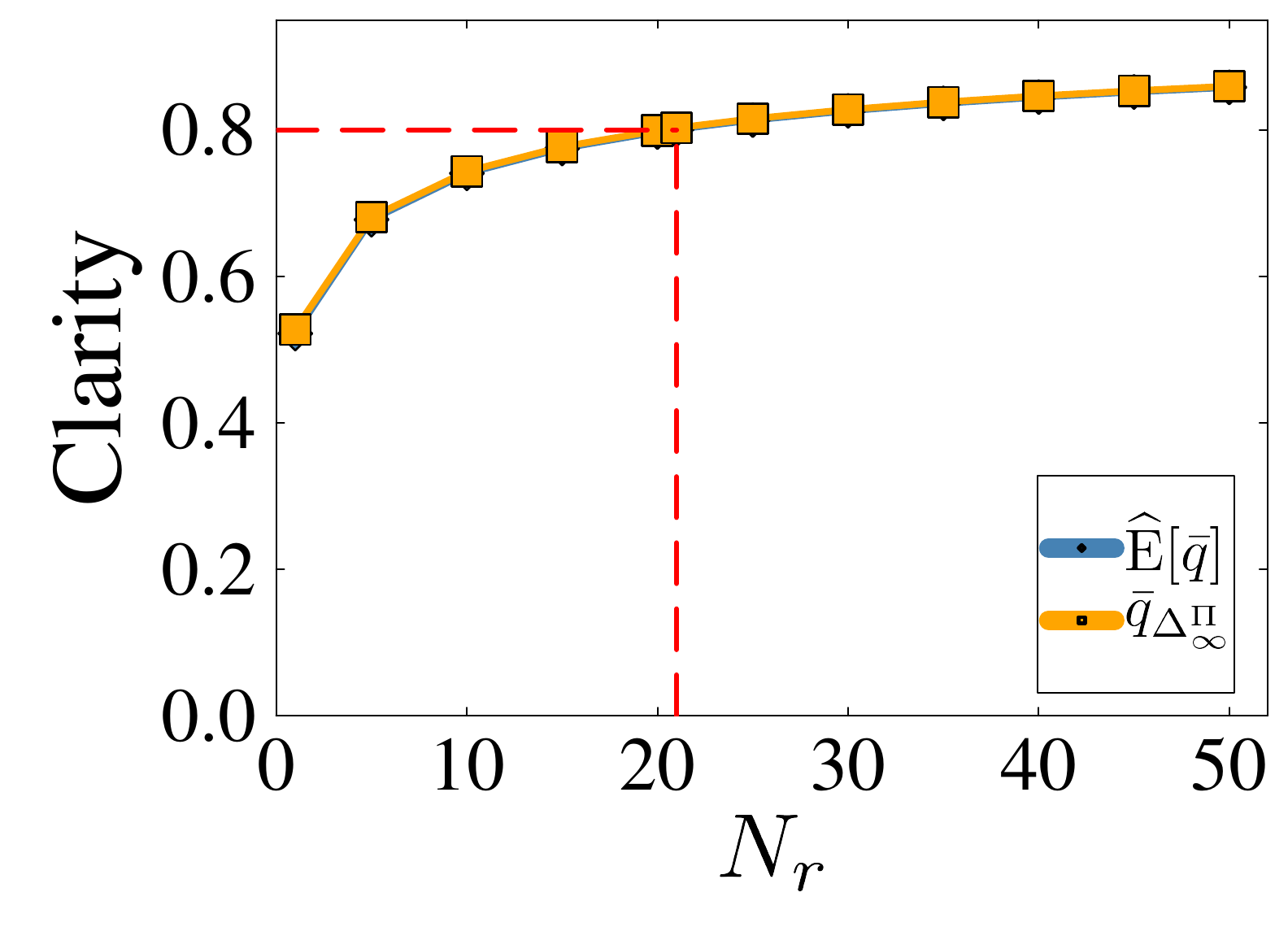}
    \hfill
    \includegraphics[width=0.52\linewidth]{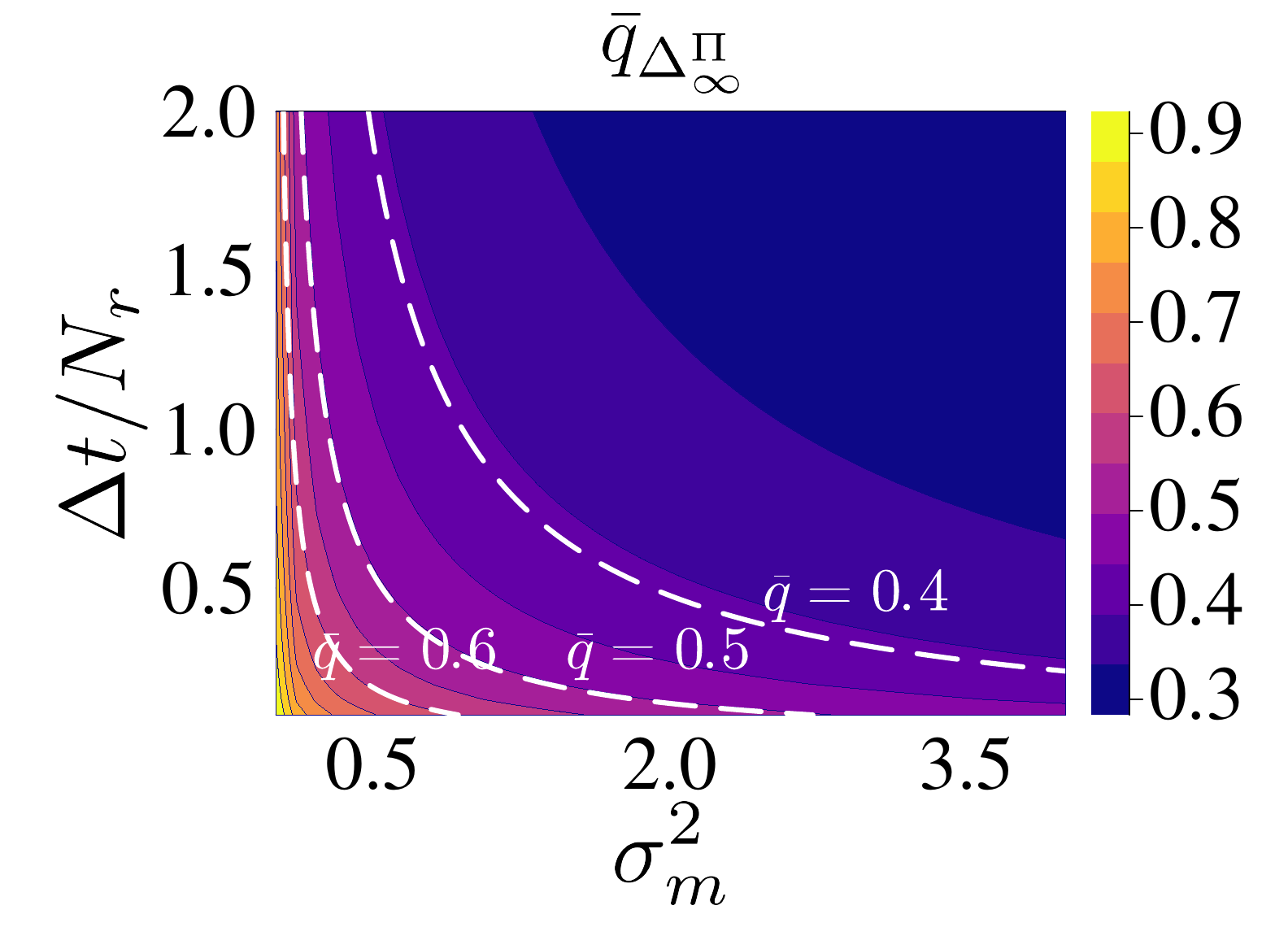}\\[-1mm]
    \makebox[0.5\linewidth]{\textnormal{(a)}}
    \hfill
    \makebox[0.49\linewidth]{\textnormal{(b)}}
    \caption{\begin{editone}Sensor-network design results. (a) The estimate $\widehat{\E}[\bar q]$ and the steady-state lower bound $\bar{q}_{\Delta^{\Pi}_\infty}$ versus the number of sensors. (b) Steady-state continuous-time lower bound on expected spatially averaged clarity $\bar{q}_{\Delta^{\Pi}_\infty}$ as a function of the measurement noise variance $\sigma_m^2$ and normalized sampling interval $\Delta t/N_r$, with $N_r=1$. Dashed curves indicate constant $\theta$, equivalently constant $\sigma_c^2=\sigma_m^2\Delta t$.\end{editone}}
    \label{fig:sensor_design}
\end{figure}

Two main consequences follow. 
First, the limiting performance is intrinsic to the kernel and system dynamics rather than an artifact of grid resolution. Second, the dependence on a single parameter $\theta$ provides a clear design principle for a given environment setting: performance can be improved equivalently by adding agents, reducing measurement noise, or decreasing sampling interval. 
Moreover, the minimum sensor number required to meet a desired performance threshold can be computed directly based on Theorem~\ref{thm:clarity_converge}. 
\begin{editone}We now revisit the design question: \emph{Given the environmental parameters, what sensor number is selected by the analytical criterion for a desired uncertainty threshold?}\end{editone}

\begin{editone}Under the conditions of Theorem~\ref{thm:continuous_time_upper_bound}, Eq.~\eqref{eq:expected_clarity_large_q_delta} implies that 
% the expected spatially averaged clarity 
$\E[\bar q(t)]$ of the continuous-time model is certified by the lower bound $\bar{q}_{\Delta^{\Pi}(t)}$ obtained from the upper-bound covariance $\bm{\Delta}(t)$.\end{editone}
Enforcing clarity constraints for all $t$ is intractable.
\begin{editone}We therefore use the steady-state clarity lower bound $\bar{q}_{\Delta^{\Pi}_\infty}$ as an analytical criterion for sensor-number design. Theorem~\ref{thm:clarity_converge} establishes that this quantity converges as the spatial grid is refined. 
    % Proposition~\ref{prop:stgpkf_continuous_limit} establishes that, over every fixed finite time horizon, the expected covariance of the discrete-time \ac{STGPKF} converges uniformly at the sampling instants to $\bm{\Delta}(t_k)$ as $\Delta t\to0$.
\end{editone}
Consequently, the agent-number design problem can be formulated as the following optimization:
\eqn{
\label{eq:sensor_number_search}
\min\; N_r
\quad\text{s.t.}\;
\bar{q}_{\Delta^{\Pi}_\infty} \ge q_{\text{target}}.
}
Here, $q_{\text{target}}$ denotes the user-defined clarity level.

\begin{editone}
For the discrete-time \ac{STGPKF},
Proposition~\ref{prop:stgpkf_continuous_limit} provides a
finite-horizon justification for the discrete-to-continuous
approximation as $\Delta t\to0$. Therefore,
Eq.~\eqref{eq:sensor_number_search} is used as a design surrogate at finite $\Delta t$, and its accuracy is evaluated numerically.
\end{editone}
\begin{editone}For a fixed set of sensing parameters, evaluating $\bar{q}_{\Delta^{\Pi}_\infty}$ from its closed-form expression requires \(O(N_g^3)\) operations and \(O(N_g^2)\) memory.\end{editone}

\begin{table}[htbp]
  \centering
  \begin{editone}
  \caption{The estimate $\widehat{\E}[\bar q]$ and $\bar{q}_{\Delta^{\Pi}_\infty}$ versus the number of sensors.}
  \label{tab:clarity_vs_nr}
  \begin{tabular}{lccc}
    \toprule
      $q_{\text{target}}$ & $N_r$ & $\widehat{\E}[\bar q]$ & $\bar{q}_{\Delta^{\Pi}_\infty}$ \\
    \midrule
   0.50  & 1  & 0.522 & 0.528 \\
   0.60  & 3 & 0.628 & 0.632 \\
   0.70  & 7 & 0.709 &  0.712 \\
   0.80  & 21 & 0.801 & 0.803\\
   0.90  & 112 & 0.899 & 0.900 \\
    \bottomrule
  \end{tabular}
  \end{editone}
\end{table}

\begin{editone}Results show that clarity increases monotonically with $N_r$, though with diminishing returns as the number of sensors grows. The selected sensor numbers satisfy the criterion in Eq.~\eqref{eq:sensor_number_search} by construction, while $\widehat{\E}[\bar q]$ closely tracks $\bar q_{\Delta^{\Pi}_\infty}$, with a small finite-sampling-interval gap.
%  Thus, Figure~\ref{fig:sensor_design}(a) and Table~\ref{tab:clarity_vs_nr} support the use of $\bar q_{\Delta^{\Pi}_\infty}$ as a practical sensor-number design surrogate.
\end{editone}

\begin{editone}With the number of sensors fixed, we can also examine how sensing capabilities affect the clarity lower bound. Figure~\ref{fig:sensor_design}(b) shows that the clarity lower bound decreases from the lower-left to the upper-right.
% ; equivalently, the covariance bound increases because the sensing intensity $\theta=N_r/(\sigma_m^2\Delta t)$ decreases as either $\sigma_m^2$ or $\Delta t$ increases. 
% For this design comparison, $\sigma_m^2$ and $\Delta t$ are varied independently, so the corresponding continuous-time noise intensity $\sigma_c^2=\sigma_m^2\Delta t$ varies across the plot. 
Each dashed curve represents a constant value of $\theta$, equivalently a constant $\sigma_c^2$ for fixed $N_r$, and therefore yields the same continuous-time \begin{editone}steady-state clarity lower bound\end{editone}. Thus, for fixed measurement noise, decreasing $\Delta t$ (increasing the measurement rate) increases $\theta$; for fixed $\Delta t$, reducing $\sigma_m^2$ has the same effect. This characterization enables explicit trade-offs between sensor noise and measurement rate prior to deployment.\end{editone}

\section{Conclusions}
\label{sec:conclusion}
This paper developed a continuous-time framework for quantifying performance bounds for multi-agent spatiotemporal estimation under stochastic sensing. We derived the first continuous-time upper bound on the expected Kalman–Bucy covariance under randomized sensing, extending discrete-time results to a broader analytical setting. Applied to spatiotemporal field estimation, the framework yields a steady-state clarity lower bound that converges to a grid-independent limit as the spatial grid is refined and is governed by a parameter combining agent number, measurement noise, and measurement rate.
Simulation results show close agreement between the continuous-time bound and the empirical discrete-time performance, supporting its use as a practical surrogate for pre-deployment sensor-number design without costly simulations.
\begin{editone}
Future work will extend the framework to mobile sensors with motion constraints, whose temporally correlated trajectories require new analysis to relate the discrete-time sensing process to the continuous-time model and may preclude the time-invariant averaged information matrix used in the current steady-state characterization. We will also consider non-uniform location sampling, for which the grid-limit and sensor-design results require a weighted spectral analysis.
\end{editone}
% Future research will consider incorporating the dynamics and motion constraints of mobile sensors into the estimation framework and address non-uniform location sampling arising from heterogeneous sensor distributions. 

% \begin{ack}                               

% \end{ack}

\bibliographystyle{plain}        % Include this if you use bibtex 
\bibliography{autosam}          
               
\appendix
\begingroup
\renewenvironment{pf}
  {
    \par\addvspace{0.4\baselineskip}%
   \noindent\textbf{PROOF.}\enspace\ignorespaces}
  {\par\addvspace{0.4\baselineskip}
}
\section{Proofs}
\subsection{Proof of Lemma~\ref{lem:concavity}}
\label{appendix:concavity}
\begin{pf}
    Let $
    \bm{\Sigma}_\beta = \beta \bm{\Sigma}_1+ (1-\beta) \bm{\Sigma}_2 
    $
    with $\beta\in[0,1]$.
    The affine part is linear, so
   \eqnN{
   &  \bm{A} \bm{\Sigma}_\beta + \bm{\Sigma}_\beta \bm{A}^\top   + \bm{Q}_c \\
   & = \beta ( \bm{A} \bm{\Sigma}_1 +  \bm{\Sigma}_1 \bm{A}^\top)
   + (1-\beta)(\bm{A} \bm{\Sigma}_2  + \bm{\Sigma}_2 \bm{A}^\top) + \bm{Q}_c.
   }
For the quadratic term,
   \eqnN{
   & \bm{\Sigma}_\beta \bm{G} \bm{\Sigma}_\beta  = (\beta \bm{\Sigma}_1 + (1-\beta)\bm{\Sigma}_2)\bm{G}(\beta \bm{\Sigma}_1 + (1-\beta)\bm{\Sigma}_2)\\
%    & =\beta^2 \bm{\Sigma}_1 \bm{G} \bm{\Sigma}_1 + (1-\beta)^2 \bm{\Sigma}_2 \bm{G} \bm{\Sigma}_2 \\
%    &\quad +\beta(1-\beta)( \bm{\Sigma}_1 \bm{G} \bm{\Sigma}_2 +  \bm{\Sigma}_2 \bm{G} \bm{\Sigma}_1)\\
%    & = \beta \bm{\Sigma}_1 \bm{G} \bm{\Sigma}_1 + (1-\beta) \bm{\Sigma}_2 \bm{G} \bm{\Sigma}_2 \\
%    &\quad -\beta(1-\beta)( \bm{\Sigma}_1 \bm{G} \bm{\Sigma}_1 +  \bm{\Sigma}_2 \bm{G} \bm{\Sigma}_2 -(\bm{\Sigma}_1 \bm{G} \bm{\Sigma}_2 +  \bm{\Sigma}_2 \bm{G} \bm{\Sigma}_1))\\
   &=\beta \bm{\Sigma}_1 \bm{G} \bm{\Sigma}_1 + (1-\beta) \bm{\Sigma}_2 \bm{G} \bm{\Sigma}_2 \\
   &\quad -\beta(1-\beta)(\bm{\Sigma}_1-\bm{\Sigma}_2)\bm{G}(\bm{\Sigma}_1-\bm{\Sigma}_2).}
   Since $\bm{G} \succeq 0$, we have $(\bm{\Sigma}_1-\bm{\Sigma}_2)\bm{G}(\bm{\Sigma}_1-\bm{\Sigma}_2) \succeq 0$.
    \eqnN{
    \bm{\Sigma}_\beta \bm{G} \bm{\Sigma}_\beta \preceq \beta \bm{\Sigma}_1 \bm{G} \bm{\Sigma}_1 + (1-\beta) \bm{\Sigma}_2 \bm{G} \bm{\Sigma}_2.
    }
    Hence, 
   \eqnN{
   % & \Ric_{G}(\bm{\Sigma}_\beta)\\
   &\Ric_{G}(\beta \bm{\Sigma}_1 + (1- \beta)\bm{\Sigma}_2) \\
   &=  \bm{A} \bm{\Sigma}_\beta +\bm{\Sigma}_\beta \bm{A}^\top   + \bm{Q}_c - \bm{\Sigma}_\beta \bm{G} \bm{\Sigma}_\beta \\
   &\succeq \beta ( \bm{A} \bm{\Sigma}_1 +\bm{\Sigma}_1  \bm{A}^\top  + \bm{Q}_c) - \beta \bm{\Sigma}_1 \bm{G} \bm{\Sigma}_1 \\
   & \quad  + (1-\beta)(\bm{A} \bm{\Sigma}_2  + \bm{\Sigma}_2A^\top  + \bm{Q}_c) - (1-\beta) \bm{\Sigma}_2 \bm{G} \bm{\Sigma}_2\\
   &= \beta \Ric_{G}(\bm{\Sigma}_1) + (1 - \beta) \Ric_{G}(\bm{\Sigma}_2)
   }
\end{pf}

\subsection{Proof of Lemma~\ref{lem:diff_of_riccati}}
\label{appendix:diff_of_riccati}
\begin{pf}
The proof is inspired by~\cite[Theorem 2.1]{freiling1996generalized}.
\eqnN{
&\Ric_{G}(\bm{\Sigma}_1) - \Ric_{G}(\bm{\Sigma}_2)\\
% &= \bm{A} \bm{\Sigma}_1 + \bm{\Sigma}_1 \bm{A}^\top  - \bm{\Sigma}_1 \bm{G} \bm{\Sigma}_1 - \bm{A} \bm{\Sigma}_2 - \bm{\Sigma}_2 \bm{A}^\top + \bm{\Sigma}_2 \bm{G} \bm{\Sigma}_2\\
&= \bm{A} (\bm{\Sigma}_1 - \bm{\Sigma}_2) + (\bm{\Sigma}_1 - \bm{\Sigma}_2) \bm{A}^\top - \bm{\Sigma}_1 \bm{G} \bm{\Sigma}_1 + \bm{\Sigma}_2 \bm{G} \bm{\Sigma}_2\\
&= \bm{A}\bm{K} + \bm{K} \bm{A}^\top - \bm{\Sigma}_1 \bm{G} \bm{\Sigma}_1 + \bm{\Sigma}_2 \bm{G} \bm{\Sigma}_2\\
&= \bm{A}\bm{K} + \bm{K} \bm{A}^\top - \bm{\Sigma}_1 \bm{G} \bm{\Sigma}_1 + (\bm{\Sigma}_1 - \bm{K}) \bm{G} (\bm{\Sigma}_1 - \bm{K})\\
&= \bm{A}\bm{K} + \bm{K} \bm{A}^\top - \bm{\Sigma}_1 \bm{G} \bm{K} - \bm{K} \bm{G} \bm{\Sigma}_1 + \bm{K} \bm{G} \bm{K}\\
% &= (\bm{A} - \bm{\Sigma}_1 \bm{G} + \frac{1}{2}  \bm{K}  \bm{G}) \bm{K} + \bm{K}(\bm{A}^\top - \bm{G} \bm{\Sigma}_1 + \frac{1}{2}  \bm{G}  \bm{K} )\\
&= (\bm{A} - \frac{1}{2} (\bm{\Sigma}_1 + \bm{\Sigma}_2)\bm{G}) \bm{K} + \bm{K}(\bm{A}^\top - \frac{1}{2} \bm{G} (\bm{\Sigma}_1 + \bm{\Sigma}_2)) \\
&= \tilde{\bm{A}} \bm{K} + \bm{K} \tilde{\bm{A}}^\top.
}
\end{pf}

\subsection{Proof of Lemma~\ref{lem:lyapunov}}
\label{appendix:lyapunov}
\begin{pf}
    Let $\bm{\Psi}(t, t_0) \in \R^{n \times n}$ be the state-transition matrix associated with $\tilde{\bm{A}}(t)$, i.e.,
    \eqnN{
    \dot{\bm{\Psi}}(t, t_0) = \tilde{\bm{A}}(t) \bm{\Psi}(t, t_0), \quad
    \bm{\Psi}(t_0, t_0) = \bm{I}.
    }    
    Define 
    $\bm{M}(t) := \bm{\Psi}(t, t_0)^{-1} \bm{K}(t) \bm{\Psi}(t, t_0)^{-T}.
    $
    Suppressing time arguments and using
    $ \frac{d}{dt} \bm{\Psi}^{-1} = -\bm{\Psi}^{-1} \tilde{\bm{A}}$ and
    $ \frac{d}{dt} \bm{\Psi}^{-\top} = -\tilde{\bm{A}}^{\top} \bm{\Psi}^{-\top}$,
    the product rule gives
    \eqnN{
    \dot{\bm{M}} &= \frac{d\bm{\Psi}^{-1}}{dt} \bm{K} \bm{\Psi}^{-T}
    + \bm{\Psi}^{-1} \dot{\bm{K}} \bm{\Psi}^{-T}
    + \bm{\Psi}^{-1} \bm{K} \frac{d\bm{\Psi}^{-T}}{dt} \\
    &= \bm{\Psi}^{-1}( \dot{\bm{K}} - \tilde{\bm{A}} \bm{K}
    - \bm{K} \tilde{\bm{A}}^\top ) \bm{\Psi}^{-T} \succeq 0
    }
    Hence, $\bm{M}(\cdot)$ is monotone nondecreasing in the Loewner order and
    $
    \bm{M}(t)\succeq \bm{M}(t_0) = \bm{K}(t_0) \succeq 0,  \quad \forall t \geq t_0.
    $
    Finally, by congruence, we have
    \eqnN{
    \bm{K}(t) = \bm{\Psi}(t, t_0) \bm{M}(t) \bm{\Psi}(t, t_0)^\top \succeq 0, \quad \forall t \geq t_0.
    }
\end{pf}

\subsection{Proof of Lemma~\ref{lem:K_cancels}}
\label{appendix:K_cancels}

\begin{pf}
    Let $\bm{S} \in \R^{N_g \times N_r}$ be the selection matrix whose $j$th column is the standard basis vector corresponding to the grid location of sensor $j$.
    We have $\bm{K}_{rr} = \bm{S}^\top \bm{K}_{gg} \bm{S}, \, \bm{K}_{rg} = \bm{S}^\top \bm{K}_{gg}, \, \bm{K}_{gr} = \bm{K}_{gg} \bm{S}.$
    Therefore,
    \eqnN{
    & \bm{K}_{rg}\bm{K}^{-1}_{gg}\bm{K}_{gr}  =  (\bm{S}^\top \bm{K}_{gg})\bm{K}^{-1}_{gg} (\bm{K}_{gg} \bm{S})
    % =\bm{S}^\top \bm{I}\bm{K}_{gg} \bm{S}
    = \bm{K}_{rr}
    }
    Thus, $\bm{K}_{rr} - \bm{K}_{rg} \bm{K}_{gg}^{-1} \bm{K}_{gr} = 0.$
\end{pf}

\begin{editone}
\subsection{Proof of Proposition~\ref{prop:stgpkf_continuous_limit}}
\label{appendix:stgpkf_continuous_limit}
\begin{pf}
% We first establish that, uniformly over \(t_k\leq T\), the discrete-time covariance \(\bm\Sigma_k\) converges in probability to the continuous-time solution \(\bm\Delta(t)\) evaluated at \(t=t_k\).
Under the i.i.d. sampling model specified above, $\bm{\Sigma}_k$ depends only on the sensing configurations before $t_k$ and is therefore independent of the newly sampled sensing matrix $\bm{H}_k$. For grid-point sensing, Lemma~\ref{lem:K_cancels} and $\sigma_m^2=\sigma_c^2/\Delta t$ give $\bm{V}_k=(\sigma_c^2/\Delta t)\bm{I}_{N_r}$. Since $\bm V_c(t_k)=\Delta t\,\bm V_k=\sigma_c^2\bm I_{N_r}$, Eq.~\eqref{eq:continuous_information_matrix} gives
\eqnN{
\bm{G}(t_k)=\bm{H}_k^\top(\sigma_c^2\bm{I}_{N_r})^{-1}\bm{H}_k,
\, \E[\bm{G}(t_k)\mid\mathcal{P}_{t_k^-}]=\bar{\bm{G}}.
}
Because the spatial grid is fixed and finite, $\bm{G}(t_k)$ is uniformly bounded. Moreover,
$
\bm{H}_k^\top\bm{V}_k^{-1}\bm{H}_k=\Delta t\,\bm{G}(t_k).
$
Thus, the prior-covariance recursion is
\eqn{
\bm{\Sigma}_{k+1}
=\bm{\Phi}(\bm{\Sigma}_k^{-1}+\Delta t\, \bm{G}(t_k))^{-1}\bm{\Phi}^\top+\bm{Q}_d.
}
In both the discrete and continuous Riccati equations, the measurement contribution cannot increase the covariance.
Hence, for every finite $T$,  $\bm\Sigma_k$ and $\bm\Delta(t)$ admit a deterministic bound independent of $\Delta t$ for $t_k\leq T$ and $t\leq T$, respectively.
Substituting the first-order expansions of \(\bm\Phi\), \(\bm Q_d\), and \((\bm\Sigma_k^{-1}+\Delta t\,\bm G(t_k))^{-1}\) into the preceding recursion, we obtain, uniformly for \(t_k\leq T\), the following first-order discrete Riccati recursion~\cite[Lemma~3.2(iv)]{salgado1988connection}:
\eqnN{
\bm{\Sigma}_{k+1}
=\bm{\Sigma}_k+\Delta t\, \Ric_{\bm{G}(t_k)}(\bm{\Sigma}_k)
+O\!\left((\Delta t)^2\right).
}
% Equivalently,
% \eqnN{
% \frac{\bm\Sigma_{k+1}-\bm\Sigma_k}{\Delta t}
% =\Ric_{\bm G(t_k)}(\bm\Sigma_k)+O(\Delta t).
% }
The i.i.d. sampling and the independence of the newly sampled $\bm G(t_k)$ from $\bm\Sigma_k$ therefore give
\eqnN{
\E\!\left[
\frac{\bm\Sigma_{k+1}-\bm\Sigma_k}{\Delta t}
\,\middle|\,\mathcal P_{t_k^-}
\right]
=\Ric_{\bar{\bm G}}(\bm\Sigma_k)+O(\Delta t).
}
% After identifying symmetric matrices with finite-dimensional vectors,
Together with the finite-horizon bound above, these equations verify the conditions of \cite[Ch.~8, Sec.~8.2.1, Thm.~2.1]{kushner2003stochastic} on $[0,T]$.
%  Specifically, the bounds on $\bm\Sigma_k$ and $\bm G(t_k)$ imply that the normalized increments above are bounded, which gives (A1.1). The last conditional-expectation equation and the continuity of $\Ric_{\bar{\bm G}}$ give (A1.2). Since $\Ric_{\bar{\bm G}}$ does not depend on $k$ or $\Delta t$, (A1.3) holds. Finally, the uniform $O(\Delta t)$ bound makes the averaged error in (A1.4) vanish.
Define the error \(\bm e_k:=\bm\Sigma_k-\bm\Delta(t_k)\). Since Eq.~\eqref{eq:Delta} has the unique solution $\bm\Delta(t)$, the cited theorem implies
\(\max_{t_k\leq T}\|\bm e_k\|_F\to0\) in probability over every finite time interval.
The deterministic finite-horizon bound and Jensen's inequality then give
\(\max_{t_k\leq T}\|\E[\bm e_k]\|_F
\leq \E[\max_{t_k\leq T}\|\bm e_k\|_F]\to 0\),
proving Eq.~\eqref{eq:stgpkf_continuous_limit}.
\end{pf}

% \subsection{Proof of Corollary~\ref{cor:discrete_bound_continuous_limit}}
% \label{appendix:discrete_bound_continuous_limit}
% \begin{pf}
%  Under the same grid-point
% scaling and finite-horizon boundedness arguments used
% in the proof of Proposition~\ref{prop:stgpkf_continuous_limit}, averaging the
% deterministic recursion in Proposition~\ref{prop:discrete_upper_bound} over
% the sensing configurations yields, uniformly for $t_k\leq T$,
% \eqnN{
% \bm\Delta_{k+1}
% &=\bm\Delta_k+\Delta t\,\Ric_{\bar{\bm G}}(\bm\Delta_k)
% +O((\Delta t)^2).
% }
% A one-step Taylor expansion of the solution of Eq.~\eqref{eq:Delta} gives
% \eqnN{
% \bm\Delta(t_{k+1})
% &=\bm\Delta(t_k)+\Delta t\,\Ric_{\bar{\bm G}}(\bm\Delta(t_k))
% +O((\Delta t)^2).
% }
% Both $O((\Delta t)^2)$ remainder terms are uniform for $t_k\leq T$.
% Let $e_k:=\|\bm\Delta_k-\bm\Delta(t_k)\|_F$. The two trajectories have the
% same initial condition, so $e_0=0$. Moreover,
% $\Ric_{\bar{\bm G}}$ is Lipschitz continuous on the bounded set containing
% both trajectories. Hence, for constants $L,C>0$ independent of $k$ and
% $\Delta t$,
% \eqnN{
% e_{k+1}\leq(1+L\Delta t)e_k+C(\Delta t)^2.
% }
% This recurrence has the standard discrete Gr\"onwall form, with step size
% $\Delta t$, growth coefficient $L$, local forcing $C(\Delta t)^2$, and
% $e_0=0$. The discrete Gr\"onwall inequality therefore gives a constant
% $C_T>0$, independent of $k$ and $\Delta t$, such that
% \eqnN{
% \max_{k:\,t_k\leq T}
% \|\bm\Delta_k-\bm\Delta(t_k)\|_F
% \leq C_T\Delta t\longrightarrow0,
% }
% which proves Eq.~\eqref{eq:discrete_bound_continuous_limit}.
% \end{pf}

\end{editone}

\subsection{Proof of Lemma~\ref{lem:kgrkrg}}
\label{appendix:kgrkrg}

\begin{pf}
Since the agents are at one of the grid points, we can write $\bm{K}_{gr} = \bm{K}_{gg} \bm{S}$
where $\bm{S} \in \R^{N_g \times N_r}$ is a column selector matrix, indicating which gridpoint the agent is at. Each column of $\bm{S}$ is an element of standard basis set. Therefore, 
\eqnN{
\E[\bm{K}_{gr}  \bm{K}_{rg}] = \E[ \bm{K}_{gg} \bm{S} \bm{S}^\top \bm{K}_{gg}]
% = \bm{K}_{gg} \E[\bm{S} \bm{S}^\top]\bm{K}_{gg}
= \bm{K}_{gg} \left( \frac{N_r}{N_g} \bm{I} \right) \bm{K}_{gg}.
}
\end{pf}

\subsection{Proof of Lemma~\ref{lem:G_bar}}
\label{appendix:G_bar}

\begin{pf}
    Since the agents take measurements at the grid locations, by Lemma~\ref{lem:K_cancels}, we have 
    \begin{editone}
    \eqnN{
    \bm{V}_c(\bm{r}) &= (\Delta t) (\sigma_m^2 \bm{I}_{N_r} + \bm{K}_{rr} - \bm{K}_{rg} \bm{K}_{gg}^{-1} \bm{K}_{gr} )\\
    &=(\Delta t \sigma_m^2) \bm{I}_{N_r}=\sigma_c^2\bm{I}_{N_r}.
    }
    \end{editone}
    Moreover, by Lemma~\ref{lem:kgrkrg}, we have 
    \begin{editone}
    \eqnN{
    &\E[ \bm{H}^\top \bm{V}_c^{-1} \bm{H}] = \frac{1}{\sigma_c^2} \E [ \bm{H}^\top \bm{H}]\\
    &= \frac{1}{\sigma_c^2} \E[\bm{C}^\top \bm{K}_{gg}^{-1/2} \bm{K}_{gr} \bm{K}_{rg} \bm{K}_{gg}^{-1/2} \bm{C}  ]\\
    &= \frac{1}{\sigma_c^2} \bm{C}^\top \bm{K}_{gg}^{-1/2} \E[\bm{K}_{gr} \bm{K}_{rg}] \bm{K}_{gg}^{-1/2} \bm{C}  \\
    &= \frac{1}{\sigma_c^2}\bm{C}^\top \frac{N_r}{N_g} \bm{K}_{gg}\bm{C}= \frac{N_r}{\sigma_c^2}\bm{C}^\top \frac{\bm{K}_{gg}}{N_g} \bm{C}
    }
    \end{editone}
\end{pf}

\par\removelastskip
\endgroup
\end{document}